\documentstyle [epsfig, emulateapj]{article}
\singlespace

 1

\def\be{\begin{equation}}
\def\ee{\end{equation}}

\def\etal{{\it et al.~}}

\def\HH{${\rm {H_2}}\,\,$}

\def\gs{\mathrel{\raise1.16pt\hbox{$>$}\kern-7.0pt
\lower3.06pt\hbox{{$\scriptstyle \sim$}}}}
\def\ls{\mathrel{\raise1.16pt\hbox{$<$}\kern-7.0pt
\lower3.06pt\hbox{{$\scriptstyle \sim$}}}}
\def\gtsima{$\; \buildrel > \over \sim \;$}
\def\ltsima{$\; \buildrel < \over \sim \;$}
\def\prosima{$\; \buildrel \propto \over \sim \;$}
\def\gsim{\lower.5ex\hbox{\gtsima}}
\def\lsim{\lower.5ex\hbox{\ltsima}}
\def\simgt{\lower.5ex\hbox{\gtsima}}
\def\simlt{\lower.5ex\hbox{\ltsima}}
\def\simpr{\lower.5ex\hbox{\prosima}}

\def\pp{\noindent\parshape 2 0truecm 17truecm 2truecm 15truecm}
\def\rf#1;#2;#3;#4 {\par\pp#1, #2, #3, #4. \par}

\def\pr{\ref@jnl{Phys.Rev}}     

\def\ie{{\frenchspacing\it i.e. }}

\def\href#1;#2 {{\bf #1} : {\em #2}}

\def\beq#1{\begin{equation}\label{#1}}
\def\eeq{\end{equation}}
\def\beqa#1{\begin{eqnarray}\label{#1}}
\def\eeqa{\end{eqnarray}}

\def\tento#1{\times 10^{#1}}

\def\kpc{{\rm \ kpc}}

\def\HH{H$_2$ }
\def\H2p{H$_2^+$ }

\def\mH2p{H_2^+}



\begin{document}
\thispagestyle{empty}
\title{INHOMOGENEOUS REIONIZATION REGULATED BY RADIATIVE AND STELLAR FEEDBACKS}
 
\author {Benedetta Ciardi\altaffilmark{1}, Andrea Ferrara\altaffilmark{2},
Fabio Governato\altaffilmark{3} and Adrian Jenkins\altaffilmark{4}}

\altaffiltext{1}{Universit\`a degli studi di Firenze, Dipartimento di
                    Astronomia, L.go E. Fermi 5, Firenze, Italy}
\altaffiltext{2}{Osservatorio Astrofisico di Arcetri, L.go E. Fermi 5, 
	  Firenze, Italy}
\altaffiltext{3}{Osservatorio Astronomico di Brera,Via Bianchi 46, 23807 Merate, Italy}
\altaffiltext{4}{Physics Department, Science Labs, South Road, Durham, DH1 3LE, UK}

\slugcomment{submitted to MNRAS}
\received{---------------}
\accepted{---------------}

\begin{abstract}
We study the inhomogeneous reionization in a critical density CDM universe 
due to stellar sources, including Population III objects. The spatial distribution 
of the sources is obtained from high resolution numerical N-body simulations. 
We calculate the source properties taking into account a self-consistent 
treatment of both radiative (\ie ionizing and \HH-photodissociating photons) 
and stellar (\ie SN explosions) feedbacks regulated by massive stars. 
This allows us to describe the topology of the ionized and dissociated 
regions at various cosmic epochs and derive the evolution of H, He, and \HH 
filling factors, soft UV background, cosmic star formation rate and the final 
fate of ionizing objects. The main results are: (i) galaxies reionize the IGM 
by $z$$\approx$10 (with some uncertainty related to the gas clumping factor), 
whereas \HH is completely dissociated already
by $z$$\approx$25; (ii) reionization is 
mostly due to the relatively massive objects which collapse via H line 
cooling, while objects whose formation relies on H$_2$ cooling alone are insufficient to this aim; (iii) the diffuse soft 
UV background is the major source of  radiative feedback effects for $z\le 
15$; at higher $z$ direct flux from neighboring objects dominates;
(iv) the match of the calculated cosmic star formation history with the one 
observed at lower redshifts suggests that the conversion efficiency of 
baryons into stars is $\approx$1\%; (v) we find that a very large 
population of dark objects which failed to form stars is present by 
$z$$\approx$8. We discuss and compare our results with similar previous 
studies.    
\end{abstract}

\section{INTRODUCTION}
\label{intr}

At $z$$\approx$1100 the intergalactic medium (IGM) is expected to
recombine and remain neutral until the first sources of ionizing
radiation form and reionize it. The application of the Gunn-Peterson
(1965) test to QSOs absorption spectra suggests that the HI reionization
is complete by $z$$\approx$5.
Recently, Ly$\alpha$ emitters have been seen up to a redshift of
$z$=6.68 (Chen, Lanzetta \& Pascarelle 1999). This indicates that the
IGM was already ionized by then, but      the epoch of complete 
reionization has yet to be firmly established.  Several
authors (Shapiro, Giroux \& Babul  1994; Madau, Haardt \& Rees 1998 and 
references therein) have claimed that the known population of quasars
(showing a pronounced cutoff at $z\simgt 3$ [Warren, Hewett \& Osmer 1994; Schmidt,
Schneider \& Gunn 1995; Shaver \etal 1996]) and galaxies
provides $\approx$10 times fewer ionizing photons than are
necessary to keep  the observed IGM ionization level. Thus, additional 
sources of ionizing photons are
required at high redshift. Although
models have been proposed in which the UV photons responsible for the
IGM reionization may be emitted by a cosmological distribution of
decaying dark matter particles, such as neutrinos (see for example
Scott, Rees \& Sciama 1991), the most promising sources are early
galaxies and quasars. 

Recent observational evidence   suggest the existence of an early
population of pregalactic objects which could have contributed to the
reionization and metal enrichment of the IGM. 
Metals have been clearly detected in Ly$\alpha$ forest clouds with 
column densities low enough to be identified with a truly diffuse IGM (Cowie
\etal 1995; Tytler \etal 1995; Lu \etal 1998; Cowie \& Songaila 1998),
although discrepant metallicities have been inferred.
Naively, one can estimate that the amount of heavy elements associated with 
the number of photons required
to ionize every baryon in the universe a few times corresponds
to an IGM metallicity $Z\approx 3\times 10^{-3} Z_\odot$. This figure 
is roughly consistent with the value obtained by Cowie \& Songaila (1998),
but ten times larger than the one derived by Lu \etal (1998). The reasons
for this  discrepancy are not yet clear.
Whatever the correct level of metallicity is, these heavy elements must have
been produced by objects hosting the sites of the earliest
star formation activity in the universe. The presence of a soft, stellar
component of the UV background deduced from studies of the [Si/C]
abundance ratios in low column density absorption systems (Savaglio
\etal 1997; Giroux \& Shull 1997), also supports this view.
The question remains concerning the transport mechanism
from the production regions, which are presumably associated with high 
peaks of the density fluctuation field, into the very diffuse medium probed by 
absorption line experiments. It is currently unclear if the mechanical energy
input associated with the same massive stars producing the metals (\ie supernovae)
is sufficient to drive the metals far enough from their production site or if
additional mechanisms need be invoked. 
Although
recently Gnedin (1998), using high resolution cosmological simulations
which include a coarse modelling of a two-phase interstellar medium, has
suggested that merging is the dominant mechanism for transporting heavy elements
from primeval galaxies into the IGM, direct
ejection of the metal enriched interstellar gas by supernovae or stellar winds 
may still play an important role.

A contribution to the ionizing flux at high redshift
from an early population of quasars may provide the hard component 
required for HeII reionization.
High resolution spectra of a quasar at $z$$\approx$3 have shown
large fluctuations of the HI to HeII optical depth ratio (Reimers
\etal 1997), interpreted as evidence for a patchy HeII ionization in the
IGM. However, it is unclear if these fluctuations could
be rather caused by statistical fluctuations of the IGM density or of the
ionizing background flux (Miralda-Escud\'e 1998; Miralda-Escud\'e,
Haehnelt \& Rees 1998). Analytical results from Haiman \& Loeb
(1998b) show that HeII reionization due to sources with quasar-like spectra
should occur at approximately the same redshift as hydrogen reionization; if
the predicted HeII reionization redshift will be 
confirmed by future measurements,
this would severely constrain the number of high redshift quasars
predicted by theory. 

The study of the IGM reionization due to these primeval sources
has been tackled by several authors,
both via semi-analytical (Shapiro 1986;
Fukugita \& Kawasaki 1994; Tegmark, Silk \&
Blanchard 1994; Haiman \& Loeb 1998b; Miralda-Escud\'e, Haehnelt \& Rees
1998; Valageas \& Silk 1999, VS) and numerical (Gnedin \& Ostriker
1997, GO; Norman, Paschos \& Abel 1998; Razoumov \& Scott 1998) approaches.
One of the main difficulties of the problem is the
treatment of cosmological radiative transfer. Only recently
have numerical approaches have been attempted (Abel, Norman \& Madau 1998;
Norman, Paschos \& Abel 1998),
and these still require severe approximations to the radiative 
transfer equation. Clearly, a numerical approach offers the advantage 
of a more detailed treatment of features such as the IGM clumpiness (GO;
Shapiro, Raga \& Mellema 1998),
or the source spatial distribution. Conversely,
 semi-analytical approaches are more flexible and allow a
more thorough exploration of the parameter space.

A proper treatment of the reionization problem must take into account
 the interplay of galaxy formation with the
reionization process itself. Basically, two types of feedback, 
radiative and stellar, can be at work.
In the standard cosmological hierarchical scenario for structure
formation, the objects which form first are predicted to have masses
corresponding to virial temperatures $T_{vir}<10^4$ K. Once the gas has
virialized in the potential wells of dark matter halos, additional
cooling is required to further collapse the gas and form
stars. For a gas of primordial composition at such low temperatures the
main coolant is molecular hydrogen (Peebles \& Dicke 1968; Shapiro 1992;
Haiman, Rees  \& Loeb 1996; Abel \etal 1997a; Tegmark \etal 1997; Ferrara 1998).
We define Pop~III objects as those for which \HH cooling is required
for collapse.
After a H$_2$ molecule gets rotationally or vibrationally excited
through a collision with an H atom or another H$_2$ molecule,  a     
radiative de-excitation leads to
cooling of the gas. This mechanism has been studied in great detail by
several authors (Lepp \& Shull 1984; Hollenbach \& McKee 1989; Martin,
Schwarz \& Mandy 1996; Galli \& Palla 1998) and has been applied to the 
study of the formation of primordial small mass objects (Haiman, Thoul
\& Loeb 1996; Tegmark \etal 1997; Abel \etal 1997b). 
Primordial H$_2$ forms with a fractional abundance of $\approx 10^{-7}$
at redshifts $\gsim 400$ via the H$_2^+$ formation channel. At redshifts
$\lsim 110$, when the Cosmic Microwave Background radiation (CMB) intensity
becomes weak enough to allow for significant formation of H$^-$ ions, a
primordial fraction of $f_{H_{2}}\approx 2\tento{-6}$ (Shapiro 1992;
Anninos \& Norman 1996) is produced for model universes that satisfy the
standard primordial nucleosynthesis constrain $\Omega_b h^2 = 0.0125$
(Copi, Schramm \& Turner 1995), where $\Omega_b$ is the baryon density parameter
and $H_0= 100 h$~km~s$^{-1}$~Mpc$^{-1}$ is the Hubble constant.
This primordial fraction is usually lower than the one required for the
formation of Pop~III objects, but during the collapse phase,
the molecular hydrogen content can reach high enough values to
trigger star formation.
On the other hand, objects with virial temperatures (or masses) above
that required for the hydrogen Ly$\alpha$
line cooling to be efficient, do not rely on  H$_2$
cooling to ignite internal star formation. As the first stars form,
their photons in the energy range 11.26-13.6 eV are able to
penetrate the gas and photodissociate H$_{2}$ molecules both in the IGM
and in the nearest collapsing structures, if they can propagate that far
from their source. Thus, the existence of a UV flux
 below the Lyman limit due to primordial objects, capable
of dissociating the H$_{2}$, could strongly influence subsequent
small structure formation.
Haiman, Rees \& Loeb (1997, HRL), for example, have argued that Pop~III
objects could depress the H$_{2}$ abundance in neighbor collapsing
clouds,
due to their UV photodissociating radiation, thus
inhibiting subsequent formation of small mass structures. On the other
hand, Ciardi, Ferrara \& Abel (1999, CFA) have shown that
the ``soft-UV background'' (SUVB) produced by Pop~IIIs is well below 
the threshold required for negative feedback
to be effective earlier than $z$$\approx$20. 
In principle, the collapse of larger mass
objects can also be influenced by an ionizing background,       as
gas in halos with a circular velocity lower than the sound speed of 
ionized gas    may be prevented from collapsing due to
pressure support in the gravitational potential. Thoul \&
Weinberg (1996) (see also Babul \& Rees 1992) have however 
shown that the collapse is only delayed by this process.
We will refer to this complex network of processes as ``radiative
feedback''.

The other feedback mechanism is related to massive stars and for
this reason we will call it ``stellar feedback''.
Once star formation begins in the central regions 
of collapsed objects
it may strongly influence their evolution via the effects of mass and energy
deposition due to massive stars through winds and supernova explosions.
These processes may induce two essentially different phenomena.
Low mass objects are characterized by shallow potential wells in which
the baryons are only loosely bound and a relatively small energy 
injection may be
sufficient to expel the entire gas content back into the IGM, \ie a {\it
blowaway}, thus quenching star formation. 
Larger objects, may instead be able to at least partially
retain their baryons, although a substantial fraction of the latter
are lost in an outflow, \ie a {\it blowout} (Ciardi \& Ferrara 1997; Mac Low \&
Ferrara 1999, MF; Ferrara \& Tolstoy 1999, FT).
However, even in this case the outflow induces a decrease of the star
formation rate due to the global heating and loss of the galactic ISM.
Omukai \& Nishi (1999) have pointed out that the ionizing
radiation of the first stars formed in Pop~III objects can also produce
an abrupt interruption of the star formation by dissociating the internal
\HH content. The effect of this feedback is very similar to the blowaway, as
both processes are regulated by    massive stars. 

Our aim here is to study and describe in detail the reionization 
process of the IGM. As the ionizing sources are not spatially     
homogeneously distributed, the pattern of the ionized regions is
not uniform, \ie inhomogeneous reionization takes place. 
Thus, a         crucial ingredient of such calculations is
the realistic description of the ionized region topology in the universe. 
This can be achieved only by numerical simulations 
which track the formation and merging of dark matter halos. For our
purposes, these simulations must reach the highest possible 
mass resolution in order to unambiguously identify the very small 
Pop~III objects initiating the reionization process. 
The second fundamental ingredient is a proper treatment of radiative
and stellar feedback processes on which a great deal of previous
work, either by ourselves or other groups, has been accumulating ranging 
from semi-analytical studies,
to numerical models and hydrodynamical simulations,
particularly with regard to stellar feedback. 
Improving on such results we have been able to effectively model and
include in the computations  
a wide number of effects governed by ionizing/dissociating
radiation and massive star energy injection which will be 
presented and discussed in detail.            
The delicate link between the properties of dark matter halos,
which can be reliably derived from N-body simulations,
and those describing  their baryonic counterparts
is represented by the prescription used to
model   star formation.
The advantages that the proposed approach brings along should become evident
by reading the paper: it allows to derive self-consistent inhomogeneous
reionization models and to infer the properties of the early galaxies
producing it. In addition, its predictive power allows to design
specific tests that could validate the results. 

The plan of the paper is the following. In \S \ref{sim} we present the 
N-body numerical simulations and in \S \ref{star} we describe our
assumptions about star formation. Then we turn to the discussion of 
radiative (\S \ref{rf}) and stellar (\S \ref{stfeed}) feedbacks and a brief
sketch summarizing the possible galactic evolutionary tracks due 
to such processes is presented in \S \ref{stracks}. Results are given in 
\S \ref{res} and further discussed in \S \ref{disc}. A short summary 
concludes the paper.

\section{NUMERICAL SIMULATIONS AND HALO SELECTION}
\label{sim}

We have simulated structure formation within a periodic cube of
comoving length $L=2.55h^{-1}$ Mpc for a critical density cold dark
matter model ($\Omega_0$=1, $h$=0.5 with $\sigma_8$=0.6 at $z$=0). 
Our choice for the normalization ($\sigma_8$) corresponds roughly to
those inferred from the present-day cluster abundance (see, e.g., Eke,
Cole \& Frenk 1996 and Governato \etal 1999). We remind the reader that
once the present-day value
of $\sigma_8$ for the SCDM run is selected, the redshift epoch of all other
outputs with lower values of $\sigma_8$ is uniquely specified. For
example, for the case where the present-day normalization is chosen to
be $\sigma_8$=0.6, $\sigma_8$=0.3 output corresponds to the $z=1$ epoch.
It is useful to note that the rms linear overdensity of a
sphere with a mass equal to the box is 2.1/(1+z). At early times when the
amplitude of fluctuations are small for scales larger than the box size
the simulation box should be large enough to be representative of the
world model as a whole. But as the scale of non-linear fluctuations grows
there will come a time when the simulation volume will not be large enough
to provide a proper average.  By redshift 8, for example,
 the rms linear overdensity of
a sphere with the mass of the entire simulation is 0.23 and by then the mass
spectrum of dark halos in the simulation box may not match the real mass
spectrum very well -- particularly at the high mass end. 

The initial conditions for the simulations, were set up at $z$=100 and
integrated in time to $z$=8.3 (with intermediate outputs at $z$=29.6, 25.3, 22.1,
19.8, 18.0, 16.5, 15.4, 14.3, 11.4 and 10.9). We used a transfer function 
for CDM calculated
with CMBFAST and a baryonic fraction $\Omega_b$=0.06 (Copi, Schramm \&
Turner 1995). The simulation, which uses 256$^3$ particles (about $17$
million), was computed with {\em AP3M} (Pearce \&
Couchman 1997), a parallel P3M code. A cubic spline force softening of 
0.05~kpc~$h^{-1}$  Plummer equivalent was used so that the overall
structure of the halos could be resolved.
The good mass (particle mass is $\approx 5 \times 10^5 M_\odot$)
and force resolution allows us
to study in detail the evolution of all structures likely to host the
formation of primordial stars. 

In numerical simulations, halos can be identified using a variety of
schemes.  Of these, we have chosen one that is available in
the public domain:
FOF\footnote{http://www-hpcc.astro.washington.edu/tools/FOF/} (Davis
\etal 1985).  In this scheme, all particle pairs separated by
less than $b$ times the mean interparticle separation are linked
together.  Sets of mutually linked particles form groups that are then
identified as dark matter halos.  Other halo finders that are often
used in literature to find virialized halos are HOP (Eisenstein \& Hut
1998), the ``spherical overdensity algorithm '' or SO, that finds
spherically averaged halos above a given overdensity (Lacey \& Cole
1994) and the scheme recently developed by Gross \etal (1998).
In the present study, we adopted the linking length that Lacey \& Cole
(1994) deduced  to identify virialized halos with mean densities of
$\approx$200 times the critical density at the epoch under
consideration.  Halo masses (and ultimately,
our findings) do not depend on the halo finder used (FOF, HOP or SO),
as long as it selects halos including all particles down to the same
overdensity.  Systematic deviations between different algorithms are
of the order of a few per cent (see Governato \etal 1999, Lacey \&
Cole 1994). In our analysis, we
only consider halos consisting of 16 particles or more,
corresponding to a mass of $8 \times 10^6 \, M_\odot$.

We have compared the results of the numerical simulations for the dark
matter halo distribution, with the one obtained from the Press \&
Schechter (1974, PS) formalism,
so that we can use this semi--analytical expression to calculate
certain quantities, such as the soft-UV background (see \S~\ref{inc}). 
The power spectrum normalization is 
$\sigma_8$=0.6, consistent with the above numerical simulations.
Although PS slightly
overestimates the halo      number density, the match is remarkably good
over the entire range of redshifts considered. 

Finally, the correlation function for galaxies at high redshift has been recently
predicted (Governato \etal 1998) and then measured by Giavalisco \etal
(1998),
giving strong support to hierarchical models of galaxy
formation. Bright galaxies at high redshift form on top of the highest
peaks in the mass distribution, and so they are strongly biased with
respect to the underlying dark matter distribution.
In hierarchical models we expect a similar behavior for the most
massive halos hosting Pop~III stars.  This measurement would give a
strong constraint on the shape of the power spectrum at scales smaller
than 1~Mpc, very relevant for all theories of galaxy formation.  Indeed
numerical simulations have already shown a large discrepancy between
the shape of the rotation curves of CDM halos and those of real dark
matter dominated galaxies (see Moore \etal 1999a) as CDM models predict
dark matter halos with steeper density profiles than those of real
galaxies.
We have preliminarly obtained the two-point correlation function of
halos significantly contributing to the photon production
 between  redshift $\approx$20 and 8.3.  
In our SCDM model the correlation function of halos hosting Pop~III
objects has a (comoving) length scale of about 0.15~Mpc~$h^{-1}$, slowly
decreasing with redshift. This decrease is expected (Tegmark \&
Peebles 1998) as the number of halos of mass sufficient to host Pop~III objects
increases with time.                                
Feedback effects of the type discussed here are clearly strongly influencing the evolution of the
correlation function. 
We plan to present the results concerning the correlation properties of
these halos in a future communication.

\section{FORMING THE FIRST STARS}
\label{star}

Once the gas, driven by gravitational instabilities, has been virialized 
in the potential well of the parent dark matter halo, further
fragmentation of the gas and ignition of star
formation is possible only if the gas can efficiently cool and lose
pressure support. For a plasma of primordial
composition at temperature $T< 10^{4}$ K, the typical virial temperature
of the early bound structures, molecular hydrogen is the only
efficient coolant. Thus, a minimum \HH fraction 
is required for a gas cloud to be able to cool in a 
Hubble time.
As the  intergalactic relic H$_2$ abundance falls short of at least two
orders of magnitude with respect to the above value, the fate of a
virialized lump depends crucially on its ability to rapidly increase 
its \HH content during the collapse phase.
Tegmark \etal (1997) have addressed this question in great detail by
calculating the evolution of the \HH abundance for different halo masses
and initial conditions for a standard CDM cosmology. 
They conclude that if the prevailing conditions are such 
that a molecular hydrogen fraction of order of $f_{H_2}\approx 
5 \tento{-4}$ is produced, then the lump will cool, fragment and eventually
form stars. This criterion is met only by larger halos implying that for each 
virialization redshift there will exist
some critical mass, $M_{crit}$, such that protogalaxies with total mass 
$M>M_{crit}$ will be able to form stars
and those with $M<M_{crit}$ will fail (see their Fig.~6 for the
evolution of $M_{crit}$ with the virialization redshift).
In reality, even halos with masses smaller than $M_{crit}$ could
eventually collapse at a later time (Haiman \& Loeb 1997), with a 
delay increasing with decreasing baryonic mass for a fixed rms 
amplitude of the fluctuation. However, as we will see
below, these structures will be strongly affected by various feedbacks
which essentially erase their contribution to the reionization process;
hence     we  neglect this effect in our calculations. 
In the absence of additional effects that could prevent or delay
the collapse (such as the feedbacks discussed below), we can associate 
to each dark matter halo with $M > M_{crit}$, a corresponding baryonic
collapsed mass equal to $M_{b}=\Omega_{b} M$.
Throughout the paper we adopt a value of the baryon density parameter 
$\Omega_{b}=0.06$ consistently with the numerical simulations;
note that the set of cosmological parameters used here 
are the same as in Tegmark \etal (1997), thus allowing a direct use
of their results.

As the gas collapses  and the first stars form, stellar photons with energies
in the Lyman-Werner (LW) band and above the Lyman limit,
respectively, can photodissociate H$_2$ molecules and ionize H and He atoms
in the surrounding IGM. By this process a photodissociated/ionized 
region will be produced around the collapsed object, whose size 
will depend on the source emission properties. The radiation spectrum, $j(\nu)$, adopted
here is obtained from the recently revised version of the Bruzual \& Charlot (1993, BC)
spectrophotometric code, for a Salpeter initial mass function (IMF), a single
burst mode of star formation, and a metallicity $Z=10^{-2} Z_{\odot}$. 
This choice is supported by recent observational results concluding
that the IMF in nearby systems has a slope close 
the Salpeter value above a solar mass, while flattening at lower masses
(see Scalo 1998).
Usually, a time independent IMF is assumed by most studies, but 
indirect evidences for an IMF biased towards massive
stars at early times emerge (see Larson 1998 and references therein). 
Larson (1998) suggests that the IMF might  be well represented by 
a universal power-law at large masses,
flattening below a characteristic mass whose value is 
changing with time. The rationale for this conclusion is that 
the mass scale for star formation probably depends strongly on
temperature and since star-forming clouds were probably hotter 
at earlier cosmic times, the mass scale should also
have been correspondingly higher. This results in an 
increase of the relative number of high-mass stars formed,
and consequently of the number of ionizing photons.
The relevance of the Jeans mass scale for the IMF is still 
subject of lively debate and it is not clear to which
extent it might be responsible for the mass distribution.
In view of these uncertainties and for sake of simplicity,
we will adopt the Salpeter IMF throughout the paper. 

\begin{figure}[t]
\centerline{\psfig{figure=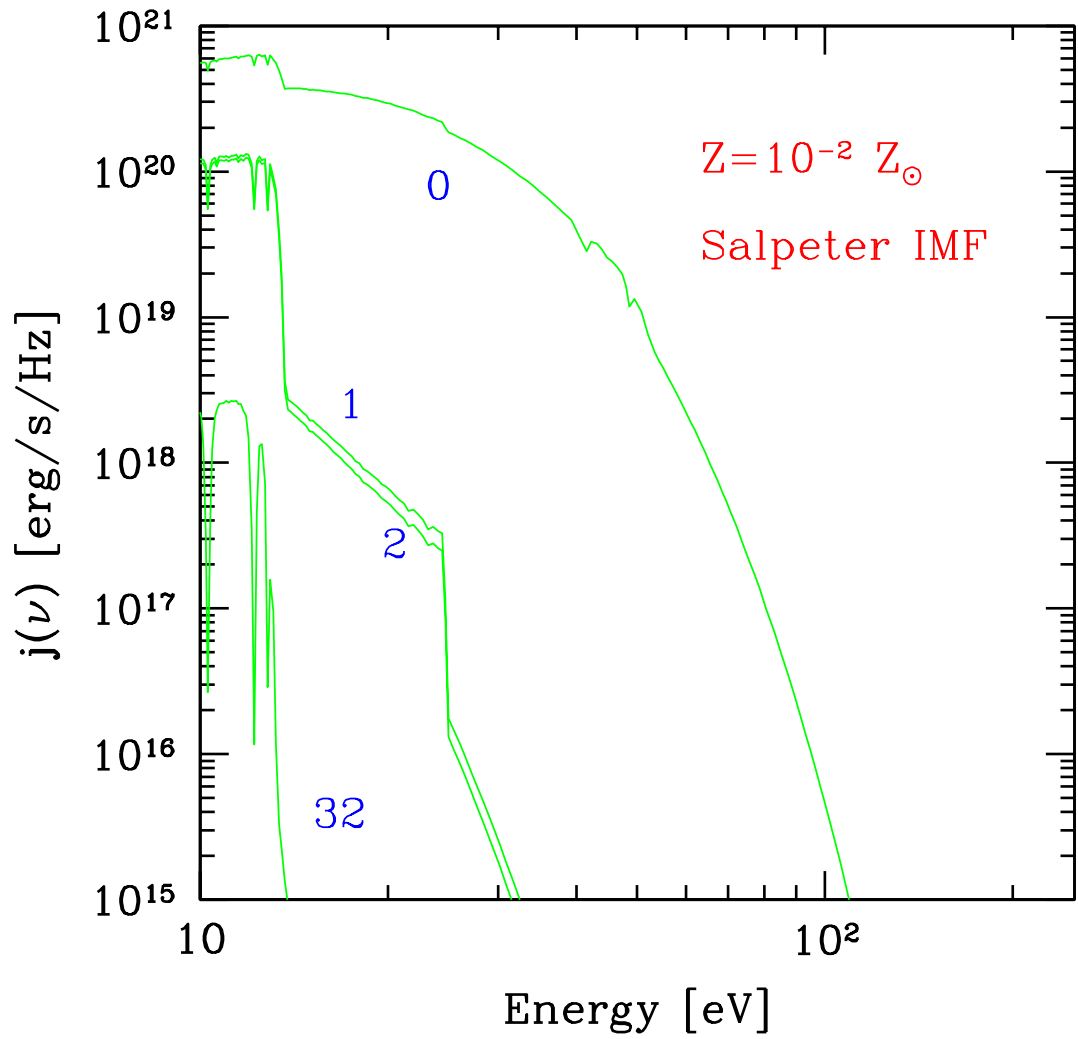}}
\caption{\label{fig1}{Evolution of the adopted emission  spectrum         
as a function of photon energy, at four different evolutionary
times $t=(0,1,2,32)\times 10^7$~yr. The specific luminosity corresponds
to one solar mass of stars formed.
}}
\end{figure}

Fig. \ref{fig1} shows the spectral energy distribution of an object per 
solar mass of stars formed at 
four different burst evolutionary times $t=(0,1,2,32)\times 10^7$~yr. 
At earlier times the spectrum is rather flat between the 13.6~eV and the HeII
edge, with a pronounced cutoff at higher energies; at later stages, it 
becomes softer with strong discontinuities both at 13.6~eV and 24.6~eV. 
The number of ionizing photons decreases considerably as the stars age.

The total luminosity of an object is determined by the mass of stars formed.
For a baryonic mass $M_b$=$10^5 \; M_{b,5}~M_\odot$, 
the corresponding stellar mass is:
\begin{equation}
M_{\star}=M_{b} f_b f_{\star} \simeq 1.2 \tento{3} \; M_{b,5} 
f_{b,8} f_{\star,15}~~M_\odot,
\label{mstar}
\end{equation}
where 
$f_b$=$0.08 \; f_{b,8}$ is the fraction of virialized 
baryons that is able to cool and
become available to form stars and $f_{\star}$=$0.15 \;
f_{\star,15}$ is the star formation efficiency. With these assumptions the 
luminosity per unit frequency at the Lyman limit, $j_0$, as obtained from the 
adopted spectrum at early evolutionary times
(Fig. \ref{fig1}), can be explicitly written in terms of $M_b$: 
\begin{eqnarray}
j_0 & = &4 \tento{20} M_{\star} \; {\rm erg} \, {\rm s}^{-1} {\rm Hz}^{-1} 
\nonumber \\
    & \simeq &4.8 \tento{23} M_{b,5} f_{b,8} f_{\star,15}\;{\rm erg} \, {\rm s}^{-1} {\rm Hz}^{-1} , 
\label{gei0}
\end{eqnarray}
of which only a fraction $j_0 f_{esc}$$\simeq$$9.6 \tento{22}$$M_{b,5}
f_{b,8} f_{\star,15} f_{esc,20}$ erg s$^{-1}$ Hz$^{-1}$ (where
$f_{esc}=0.2 \; f_{esc,20}$ is the photon escape fraction from the
proto-galaxy) is able to escape into the IGM. 
This accounts at least approximately for absorption occurring in the
host galaxy.

As $f_{b}, f_{\star}$ and $f_{esc}$ are the main parameters
involved in the calculation, aside from the
cosmological ones fixed by the simulations, it is useful to discuss them
in more detail.
Primordial stars can only form in the fraction of gas that
has been able to cool, and the cooled-to-total baryonic mass ratio
in the collapsed objects is only a few percent. Numerical simulations
(Abel \etal 1997b; Bromm, Coppi \& Larson 1999) 
have shown that $f_b$ can be $ \approx$8\%, which 
we assume as our fiducial value.
The star formation efficiency, $f_{\star}$, is rather uncertain and
dependent over the properties of the star formation environment. Even
its definition is not completely unambiguous, as star formation usually takes
place in the densest cores of molecular clouds and the 
efficiency in the cores is obviously different from the one averaged on 
the parent molecular cloud. 
We define $f_{\star}$ as the fraction of cooled (probably
molecular) gas that has been converted into stars. 
The formation of a bound cluster system requires a star formation
efficiency of nearly 50\%, where the cloud disruption is sudden and
nearly 20\% where cloud disruption takes place on a longer timescale
(Margulis \& Lada 1983; Mathieu 1983).
The highest efficiency estimated for a
star-forming region is about 30\% for the Ophiuchi dark cloud (Wilking \&
Lada 1983; Lada \& Wilking 1984). 
Lada, Evans II \& Falgarone (1997) have studied the physical properties 
of molecular cloud
cores in L1630, finding efficiencies ranging from 4\% to 30\%
for different cores. Pandey, Paliwal \& Mahra (1990) have investigated the
effect of variations in the IMF on the star formation efficiency in clouds 
of various masses. They conclude that the efficiency is      lower  
if the most massive, and consequently most destructive, stars form earlier.
On a larger scale, Planesas, Colina \& Perez-Olea (1997)
have studied the molecular gas properties and star formation in nearby
nuclear starburst galaxies, showing the existence of giant molecular
clouds ($M_{H_2} \approx 10^8-10^9 M_{\odot}$) with associated HII
regions where the star formation process is characterized by being short
lived ($< 3 \times 10^7$ yr) and with an overall gas to stars conversion
less than 10\% of the gas mass. Giannakopoulou-Creighton, Fich
\& Wilson (1999) have recently determined the star formation efficiency
in two M101 giant molecular clouds, finding the values 6\% and $>$11 \%.
In spite of great efforts to derive such an important quantity, 
various authors obtain results uncomfortably different, even when 
studying the same star forming complex (see the prototypical case of
30 Dor). Given this situation, we use the educated guess $f_{\star}=15\%$.
Finally, the value $f_{esc}$=0.2 is an upper limit derived from
observational (Leitherer \etal 1995; Hurwitz, Jelinsky \& Dixon 1997) and
theoretical (Dove \& Shull 1994; Dove, Shull \& Ferrara 1999) studies. 
The latter authors, in particular, have included the effects of 
superbubbles in the computation of $f_{esc}$, finding a value 
$f_{esc} \approx 6\%$ for the case of coeval star formation history,
which should be relevant to the present case. These studies 
concentrate on large disk galaxies like the Milky Way and it is not 
clear to which extent they can be applied to the small objects
populating the early universe; also, the much lower dust-to-gas ratio 
could make the escaping probability higher. Again, $f_{esc}=0.2$ can 
only be seen as a reference value.  
It is worth noting that while for all processes considered here only
the product $f_b\times f_{\star}$ enters  
the model, $f_{esc}$ is instead an independent parameter; 
this reduces the number of effective free parameters to two.

The total ionizing photon rate, $S_i(0)$, from the formed stellar cluster
can be  written as:  
\begin{equation}
S_i(0)=\int_{\nu_l}^{\nu_u} \frac{j(\nu)}{h \nu} d\nu,
\label{si01}
\end{equation}
where $\nu_l$=13.6 eV and $\nu_u$=150 eV. As seen from Fig.~\ref{fig1} 
the stellar spectrum drops off sharply above 100 eV even 
shortly after the burst; the choice of the upper integration limit
follows from this remark.
For the spectrum in Fig.~\ref{fig1} at $t=0$, eq.~(\ref{si01}) becomes:
\begin{eqnarray}
S_i(0) & = & 5.69 \tento{-1} (j_0/h_P) f_{esc} \nonumber \\
       & \simeq &8.18 \tento{48} \, 
       M_{b,5} f_{b,8} f_{\star,15} f_{esc,20} \; {\rm s}^{-1};
\label{si0}
\end{eqnarray}
analogously we can define the LW  photon production rate, $S_{LW}$, 
which is  typically $S_{LW}=\beta S_i(0)$, with $\beta \approx 1.5$
at early  evolutionary times.
The UV photons create a cosmological HII region in the surrounding IGM,
whose radius, $R_{i}$, 
can be approximated by the Str\"omgren (proper) radius, $R_{s}=[3
S_{i}(0)/(4 \pi n_{H}^{2} \alpha^{(2)})]^{1/3}$, where $n_{H}= 8\times
10^{-6} \Omega_b h^2 (1+z)^3$~cm$^{-3}$ is the IGM hydrogen number
density and $\alpha^{(2)}$ is the hydrogen recombination rate to levels
$\ge 2$. In general, $R_{s}$
represents an upper limit for $R_{i}$, since the ionization front fills
the time-varying Str\"omgren radius only at very high redshift, $z
\approx 100$ (Shapiro \& Giroux 1987). For our reference parameters it is:
\begin{equation}
R_{i} \simlt R_{s} = 0.05 \, \left( {\Omega_b h^2} \right)^{-2/3}
(1+z)_{30}^{-2} \; S_{47}^{1/3}  \;\; \kpc,
\label{rion}
\end{equation}
where $S_{47}=S_{i}(0)/(10^{47}$ s$^{-1}$) and $(1+z)_{30}=(1+z)/30$. 
In our calculation we use the exact solution for $R_i$ as numerically
calculated in CFA, and whose analytical approximation is given by 
Shapiro \& Giroux (1987). Typically, $R_i$ is about 1.5-2.0 times
smaller than $R_s$ and cosmological expansion cannot be neglected.
Clearly, these solutions assume that the ionizing object is irradiating a 
volume of the IGM which was previously neutral. If instead the object is
located inside (or overlapping) a preexisting ionized sphere, a
detailed solution of the radiative transfer would be necessary. This occurrence
can result in an underestimate of our calculated ionized volume when 
substantial sphere overlapping takes place, \ie close to the reionization epoch.

Similarly to $R_s$,  we can define the HeII Str\"omgren radius as $R_{s,He}=[3
S_{i,He}(0)/(4\pi n_{He^{++}} n_e \alpha_{He^{++}})]^{1/3}$, where
$S_{i,He}(0)$ is the rate of ionizing photons with $h\nu >$ 54.4 eV,
$n_{He^{++}}$ and $n_e$ are the He$^{++}$ and electron number density,
respectively, and $\alpha_{He^{++}}$ is the He$^{++}$ recombination
rate. For the adopted spectrum it is:
\begin{eqnarray}
S_{i,He}(0) & = & 1.87 \tento{-3} (j_0/h_P) f_{esc} \nonumber \\
       & \simeq &2.68 \tento{46} \,
M_{b,5} f_{b,8} f_{\star,15} f_{esc,20} \; {\rm s}^{-1}.
\label{si0he}
\end{eqnarray}
We have derived $n_{He^{++}}$ and $n_e$ assuming that the helium and
hydrogen neutral fraction is negligible. For our reference parameters and
given eqs.~(\ref{si0}) and~(\ref{si0he}), the
radius of the sphere of doubly ionized helium is:
\begin{equation}
R_{i,He} \simlt R_{s,He}=6 \times 10^{-3} \, \left( {\Omega_b h^2}
\right)^{-2/3}
(1+z)_{30}^{-2} \; S_{47}^{1/3}  \;\; \kpc,
\label{rihe}
\end{equation}
thus $R_{s,He}\simeq 0.1 R_{s}$.

Given a point source that radiates $S_{LW}$ photons per
second in the LW bands, a good estimate of the maximum radius of the H$_2$
photodissociated sphere is the distance at which the
(optically thin) photo--dissociation time becomes longer than the Hubble
time:
\begin{eqnarray}
R_{d} \simlt 2.5 \, h^{-1/2} (1+z)_{30}^{-3/4} S_{LW,47}^{1/2} \;\;
\kpc,
\label{rdiss}
\end{eqnarray}
where $S_{LW,47}=S_{LW}/(10^{47}$ s$^{-1}$). 
CFA also considered the time dependent evolution of both the ionization
and dissociation fronts; here we use the exact value for $R_i$ and
the equilibrium values
given by the equations above for both $R_{i,He}$ and $R_d$.
The latter values represent upper limits to the extent of the
spheres as emphasized above: ionized regions at high redshift
fill only partially the corresponding Str\"omgren spheres, while for the
dissociated regions the radius $R_d$ is completely filled, although
on a relatively 
long time scale. For example an object born at $z=30$ will have
a fully developed surrounding dissociated sphere by $z\approx 20$.

To each dark matter halo in which stars can form
we assign the ionized and dissociated
spheres produced by its stellar cluster according to the above
prescriptions.
This allows to derive the three dimensional structure and
topology of the ionized/dissociated regions as a function of cosmic
time.

\section{RADIATIVE FEEDBACK}
\label{rf}

In addition to their local effects, the first objects will also 
produce                    UV  radiation which could 
in principle introduce long range feedbacks on nearby collapsing halos.
Particularly relevant is       the soft UV background in the LW bands, 
as by  
dissociating the H$_2$, it could influence the star formation 
history of other small objects preventing their cooling.

CFA have argued that
the SUVB produced by Pop~IIIs is below the threshold required for
the negative feedback on the subsequent galaxy
formation to be effective before $z \approx 20$. At later times
this feedback becomes important  for objects with 
$T_{vir} \simlt T_H=10000$ K, corresponding to a mass $M_H$=$4.4
\tento{9} M_{\odot} \, (1+z_{vir})^{-1.5} h^{-1}$ (Padmanabhan 1993),
where $z_{vir}$ is the redshift of virialization, for which cooling via
Ly$\alpha$ line radiation is possible.
Proto-galaxies with $M > M_H$ are not affected by the negative 
feedback and are assumed to form. In principle even for these larger objects 
a different type of feedback can be at work: halos 
with a circular velocity lower than the sound speed of the 
gas in an ionized region may be prevented from collapsing due to 
gas pressure support in the gravitational potential. Thoul \& Weinberg
(1996) (see also Babul \& Rees 1992) have shown that the collapse is 
only delayed by this
process, and therefore we neglect this complication.

At higher redshift the radiative feedback can be induced by the direct dissociating flux from a nearby object. 
In practice, two different situations can occur: i) the collapsing
object is outside the dissociated spheres produced by preexistent objects: 
then its formation could be affected only by the 
SUVB ($J_{LW,b}$), as by construction the direct flux 
($J_{LW,d}$) can only dissociate molecular hydrogen on time 
scales shorter than the Hubble time inside $R_d$;
ii) the collapsing object is located 
inside the dissociation sphere of a previously collapsed object:
the actual dissociating flux in this case is essentially given by
$J_{LW,max}=(J_{LW,b}+J_{LW,d})$.
It is thus assumed that, given a forming Pop~III, if the incident
dissociating flux ($J_{LW,b}$ in the former case, $J_{LW,max}$ in the
latter) is higher than the minimum flux required for negative
feedback ($J_{s}$), the collapse of the object
is halted. This implies the existence of a population of "dark objects"
which were not able to produce stars and, hence, light.

\subsection{Minimum flux for negative feedback}
\label{nf}

As already stated, in the absence of an external dissociating flux,
an object can form if it satisfies the condition $M > M_{crit}$. In
this subsection we investigate how this minimum mass is changed
by the requirement that the object is able to self-shield from an external
incident dissociating flux that could deplete the H$_2$ abundance and
prevent the collapse of the gas. 
To this aim, we consider  the non-equilibrium
multifrequency radiative transfer of an incident spectrum inside
a homogeneous gas layer, and study the evolution of the following 
nine species: 
H, H$^-$,
H$^+$, He, He$^+$, He$^{++}$, H$_2$, H$_2^+$ and free electrons. 
The energy equation can be written as (Ferrara \& Giallongo 1996):
\be
{k\over (\gamma-1)} {d\over dt}\left[ T \sum_X {h_X\over \mu_{X^i}}\right] = -
{1\over n}{\cal L}(X^i)+ {p\over n^2}  {dn\over dt},
\ee
where $T$, $n$ and $p$ are the gas   temperature, number
density and pressure, respectively.
The gas is assumed of primordial composition with a helium abundance
equal to $n_{He}=0.1~n_H$ and total number density $n$. 
The symbol X= H, He, \HH denotes the species considered,
and $i$ its state of ionization; obviously,
$i=0$ for hydrogen, and $i=0,1$ for helium; 
$h_X=n_X/n$ is the relative abundance of the species $X$;
$\mu_{X^i}=(i_{max}+1)^2[1+\sum_i^{i_{max}}  (i+1) x_{X^{i+1}}]^{-1}$, with
$i_{max}=0,1$ for H and He, respectively and $x_{X^{i+1}}=n_{X^{i+1}}/n_X$ 
the fractional density of the ionization state $i+1$ of the element $X$;
$\gamma$ is the specific heat ratio assumed to be constant and equal to 5/3.
The function ${\cal L}$ represents the net cooling rate per unit volume and is
given by the difference between heating and cooling:
\be
{\cal L}(X^i)=\sum_i \sum_X n(X^i)
{\cal H}(X^i) - \sum_i \sum_X  \Lambda (X^i).
\ee

The cooling term $\Lambda$ (in erg cm$^{-3}$ s$^{-1}$)
includes collisional ionization  and excitation
of H, He, He$^+$,
recombination to H, He, He$^+$, dielectronic recombination to He,
free-free (Black 1981), Compton cooling
(Peebles 1971) and H$_2$ cooling (Martin, Schwarz \& Mandy 1996); ${\cal
H}$ includes the heating terms due to photoionization of H,
He, He$^+$, H$_2$, photodissociation of H$_2^+$ and H$_2$, and
H$^-$ photodetachment and is given by:
\begin{equation}
{\cal H}(X^i)= \int^\infty_{\nu_X} d\nu \, \frac{ 4 \pi J_s(h\nu)}{h\nu}
h(\nu-\nu_X) \sigma_{X^i}(\nu),
\label{hk}
\end{equation}
where $\nu_{X}$ indicates the ionization limit for each species,
$\sigma$ is the photoionization cross-section and $J_s$ is the incident
flux, given below.
 The various cross sections are given in Abel \etal
(1997b), apart from  $\sigma_{21}$ (taken from Brown 1971) and $\sigma_{25}$ 
(Shapiro \& Kang 1987), and they are numbered according to the nomenclature of
Abel \etal (1997b). We have assumed that He, H$^-$ and H$_2^+$ are in equilibrium as all
reactions determining the H$^-$ abundance occur on much shorter time
scales than those relevant to the H$_2$ chemistry; He and H$_2^+$
do not influence substantially the final results. The chemical network 
includes the 27 reactions listed in Abel \etal (1997b). 
We have adopted the same
rates of that paper except for $k_1$ (taken from Cen 1992), $k_2$ (Black 1981),
$k_{11}$ and $k_{15}$ (Shapiro \& Kang 1987).
The photoionization rate $\gamma_p$ is given by:
\be
\gamma_p(X^i)=\int_{\nu_{X}}^{\infty} d\nu{4 \pi J_s(h\nu)\over h \nu}
\sigma(X^i) \left[1+\phi(X^i)\right],
\ee
where  $\phi$ is the secondary ionization rate.
We have adopted the ``on the spot'' approximation in which
the diffuse field photons are supposed to be absorbed close to the point
where they have been generated.

To derive generally valid results for the negative feedback we
have taken the same incident spectrum presented in Fig. \ref{fig1} but 
at a later evolutionary time ($2\times 10^7$~yr after the burst) in order 
to roughly average over the different population evolutionary stages.
An useful analytical approximation to the adopted spectrum is: 
\begin{equation}
J_{s}(E)=J_{s,0} \left\{
\begin{array}{ll}
\beta &  E        < 13.6 \; {\rm eV},\\
(E       /13.6)^{-3.5} & 13.6 \; \le E        < 25\; {\rm eV},\\
5 \tento{-3} e^{-(E       -25)/3.3} & 25 \; \le E        < 50 \; {\rm eV},\\
2.6 \tento{-6} e^{-(E       -50)/1.5} & 50 \; \le E        \le
150 \; {\rm eV},\\
\end{array}\right.
\label{gei1}
\end{equation}
with $\beta$=50 and $E       =h\nu$.
Here $J_{s,0}$ (units of erg s$^{-1}$ cm$^{-2}$ Hz$^{-1}$ sr$^{-1}$) is
the parameter with respect to which we quantify the negative feedback at
a given redshift.

Initially,             we assume that the gas layer is homogeneous, with a
mean density $\langle \rho_b \rangle$ 18 $\pi^{2}$ times higher than      
the background intergalactic medium.
The fraction of free electrons is 
taken to have the relic value after recombination $x=10^{-5}
\Omega_b^{-1} \Omega_0^{1/2} h^{-1}$ (see for example Blumenthal \etal
1984); in principle, $x$ could have changed during the
virialization process: we neglect this possibility.             
We use equilibrium values for the He ionic abundances;
H$^{-}$ and H$_2^+$ abundances can be arbitrarily small, as
their choice does not influence the results. The initial condition for
the molecular abundance is slightly more delicate; we determine it 
as follows. For a given $M_{crit}$ (the mass of an object able to collapse
in the absence of radiation) we can determine the corresponding virial 
temperature $T_{vir}$. This represents the initial condition for the gas 
temperature. The initial \HH abundance is set by the value required to 
satisfy the condition that the free-fall time of the object is equal to
its cooling time at temperature $T_{vir}$. This assures that, if no feedback
is imposed on the object, it would collapse and form stars.

\begin{figure}[t]
\centerline{\psfig{figure=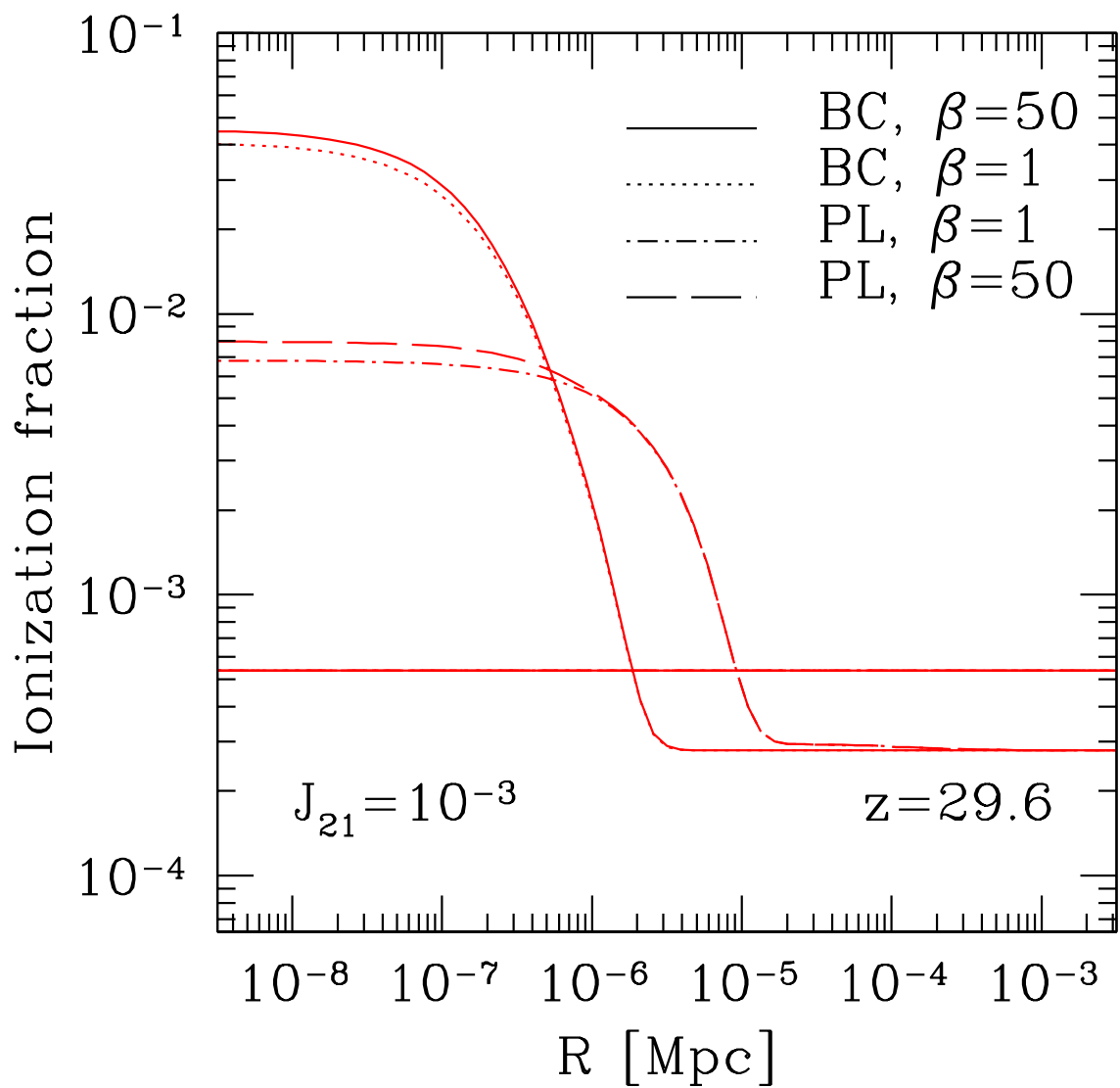}}
\caption{\label{fig2}{Evolution of the ionized atomic hydrogen fraction
as a function of depth, for a Pop~III at $z=29.6$ and an incident
flux $J_{s,0}=J_{21} 10^{-21}$ erg s$^{-1}$
cm$^{-2}$ Hz$^{-1}$ sr$^{-1}$. The curves are derived for the BC stellar  
spectrum calculated at $t=2\times 10^7$~yr
and, for comparison, for a power law spectrum (PL)
with $\alpha$=1.5. Different values of the $\beta$ parameter are
shown.   The horizontal line represents the initial condition.}}
\end{figure}

\begin{figure}[t]
\centerline{\psfig{figure=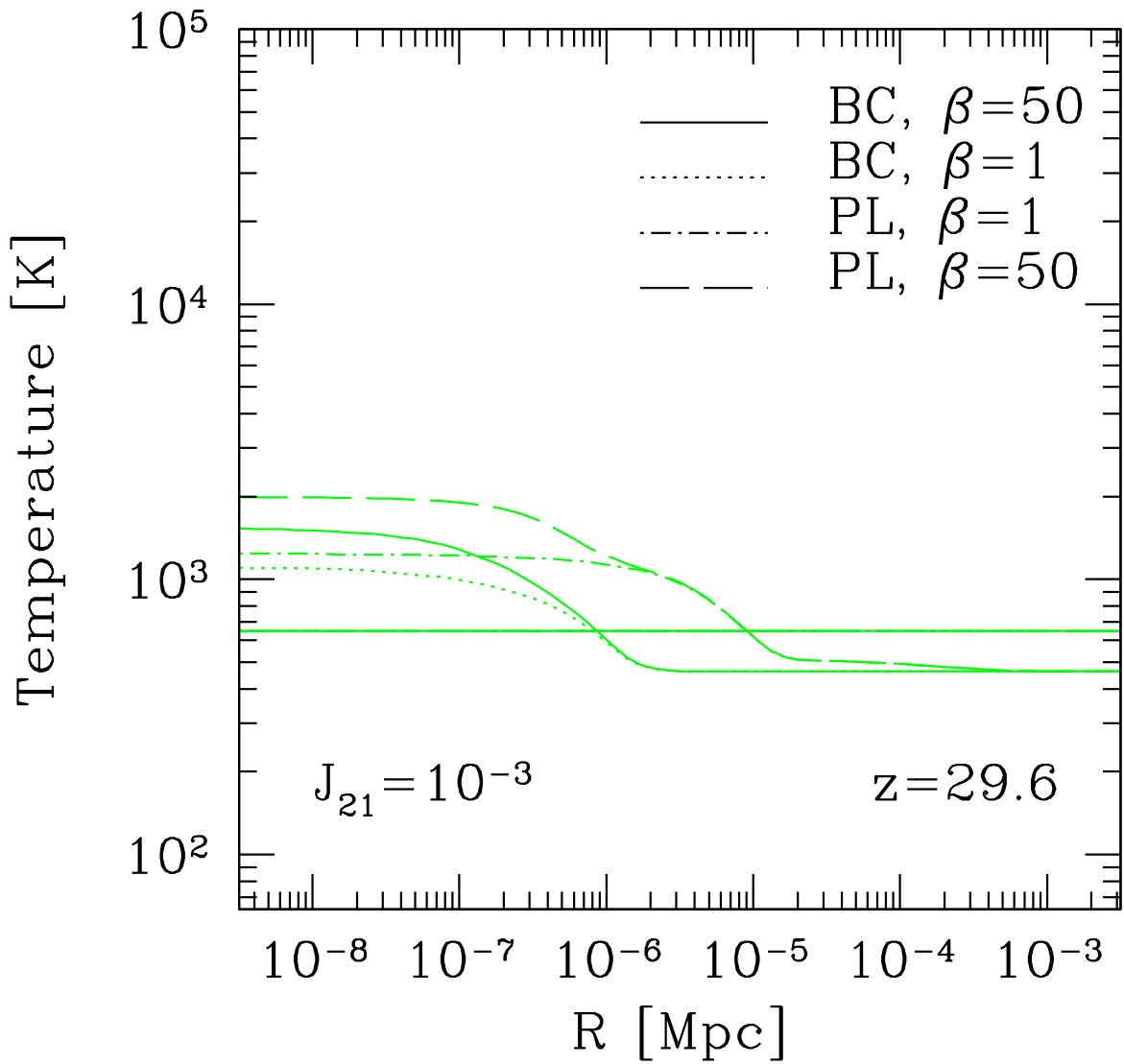}}
\caption{\label{fig3}{As Fig.~\ref{fig2} for the gas temperature.}}
\end{figure}

\begin{figure}[t]
\centerline{\psfig{figure=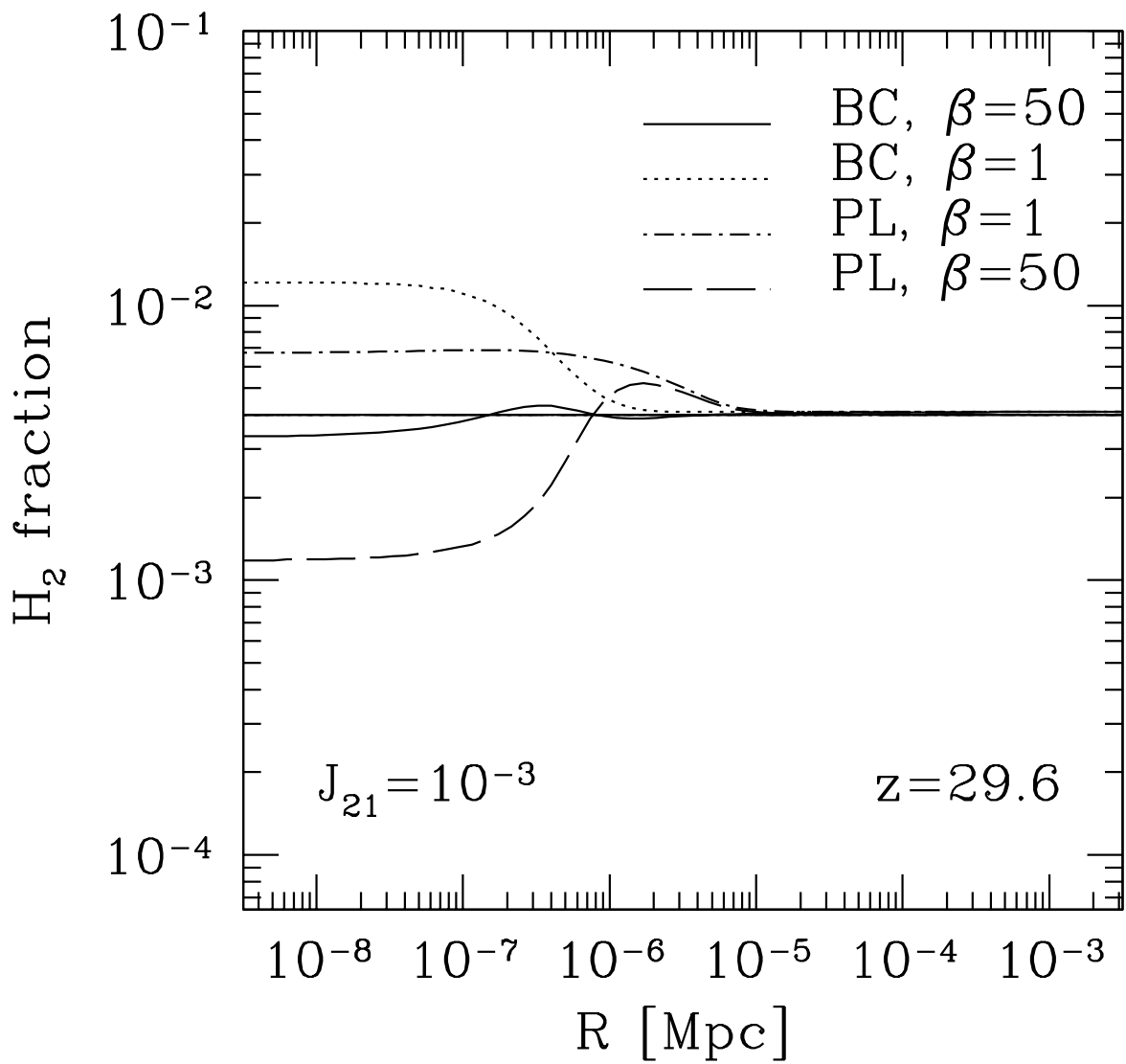}}
\caption{\label{fig4}{As Fig.~\ref{fig2} for the neutral molecular hydrogen
fraction.}}
\end{figure}

Starting from these
initial conditions we have followed the chemical evolution up to a free-fall time. 
These     
conditions are such that in the absence of radiation the 
fraction of \HH is high enough to allow for the collapse of the object on
a time scale comparable to the free-fall time. The effect of the radiation 
field is to decrease this abundance in the external regions which therefore
cannot cool.
In the interior, instead, where the \HH abundance
essentially remains equal to the initial one, the collapse can proceed unimpeded. For
this reason the collapse conditions must be checked after a free-fall time. 
As an example we discuss the case of an  
object which virializes at $z=29.6$ and to which an incident flux
with spectrum given by eq.~(\ref{gei1}) with $J_{s,0}=J_{21}
10^{-21}$ erg s$^{-1}$ cm$^{-2}$ Hz$^{-1}$ sr$^{-1}$ has been imposed. 
The critical mass for collapse, $M_{crit}$, corresponding to that
redshift is $M_{crit}=9 \times 10^5 \, M_\odot$; this fixes the initial
H$_2$ fraction as explained above.
Figs.~\ref{fig2}-\ref{fig4} show the corresponding hydrogen ionization 
fraction, $x$, the gas temperature, T, and the neutral molecular
hydrogen fraction, $f_{H_2}$, after a free-fall time.
For sake of comparison, we also show the cases relative to 
a power law spectrum (PL) with 
$\alpha$=1.5 and a cutoff energy of 40 keV. 
Different values of the $\beta$ parameter, describing the
relative number of photodissociating and ionizing photons in the
spectrum, are studied in order to cover a wide range of evolutionary 
phases of the stellar energy emission. As seen from Fig.~\ref{fig1},
the value of $\beta$ tends to increase as the stellar cluster ages.
Our standard case is the BC spectrum with $\beta$=50, appropriate to an
age of $2\times 10^7$~yr.

From Fig.~\ref{fig2}, it is clear that the
ionization fraction is mostly determined by photons with energies above the
Lyman limit; however, in the cases with $\beta$=1, as the final abundance 
of \HH is increased with respect to the initial one (see
Fig.~\ref{fig4}), electrons are depleted by the
$H^-$ molecular hydrogen formation channel, with respect to the cases
with $\beta$=50. The 
high energy photons present in the power-law spectrum penetrate 
deeper inside the layer and consequently the initial ionization is enhanced
over a larger depth than for      
the BC spectrum, although with a
lower ionization level; this feature is typical of hard ionizing spectra. 
For the same reason, a PL
spectrum heats the cloud to higher temperatures and
the plateau extends into deeper regions (Fig.~\ref{fig3}). Lower temperatures
and higher $f_{H_2}$ are obtained for smaller values of $\beta$
(see Fig.~\ref{fig3}),
when the H$_2$ cooling is more efficient. It is clear from
Fig.~\ref{fig3}, that, although in the external regions of the layer
heating is dominant due to photoelectric effect, in the inner regions,
where the medium becomes optically thick to LW photons, after a free-fall time the
temperature has decreased by about 30\% due to 
H$_2$ cooling. In Fig.~\ref{fig4} the $f_{H_2}$ profile
is shown. In the external regions, 
the most important mechanisms for the  H$_2$ destruction 
are direct photoionization by photons with energies above 15.4 eV and
LW photons dissociation. 
A PL spectrum with the same value of $\beta$ yields a larger
\HH destruction,  while if the upper PL spectrum cutoff is increased
the curves shift towards larger depths due to penetrating photons.
Note that when $\beta$=1, H$_2$ is formed rather than destroyed
due to the larger abundance of free electrons and paucity of LW
photons available. Finally, the $f_{H_2}$ smoothly approaches the
initial value in the internal regions due to
the loss of free electrons (see Fig.~\ref{fig2}). 

We would like to briefly comment on the differences between the above
curves for the \HH abundance and the analogous ones derived by HRL. 
These are mainly due to: (i) a different incident flux and 
(ii) their assumption of chemical and ionization equilibrium at a 
constant temperature, which could not be reached within a free-fall time. 
As they also note in a previous paper (Haiman, Rees \& Loeb 1996), 
this might lead to an overestimate of the \HH abundance.
In addition, they 
used the Lepp \& Shull (1984) cooling function, that, at the low
densities of interest here, is higher than the one we have adopted 
(for a comparison see Galli \& Palla 1998). 

As shown above, after a free-fall time the temperature in the inner 
regions of the protogalactic cloud 
has considerably decreased due to \HH cooling.
It is normally assumed that this process would eventually lead to
the collapse, fragmentation and star formation. These regions
are therefore  not affected by the imposed external flux
as they can efficiently self-shield.
The gas in the outer regions, up to the shielding radius, $R_{sh}(J_{s,0})$,
is instead globally heated by 
radiation and fails to collapse. On this basis, we can then define
the minimum total mass required for an object to self-shield from 
an external flux of  intensity $J_{s,0}$ at the Lyman limit, as
$M_{sh}=(4/3) \, \pi \langle\rho_{h}\rangle R_{sh}^3$, where 
$\langle\rho_{h}\rangle=\langle \rho_b\rangle \Omega_b^{-1}$ is 
the mean dark halo matter density. 
In practice,  $R_{sh}$ is defined as the point where the temperature profile
has increased by 0.01\% from the inner flat curve (see Fig.~\ref{fig3}).
Values of $M_{sh}$ for different values of $J_{s,0}$ have 
been obtained at  various redshifts and will be then used in
the calculations presented below; these curves are shown in 
Fig.~\ref{fig5}. We point out that the results depend only very 
weakly on the detailed shape of the spectrum as long as this is
produced by stellar sources with a relatively soft spectrum; 
differences up to a factor of several can be found if instead a   
hard spectral component is present (Haiman, Abel \& Rees 1999). 
In this paper we restrict our analysis to the first case. 
Protogalaxies with masses above $M_H$ for    which
cooling is predominantly contributed by Ly$\alpha$ line cooling 
will not be affected by the negative feedback studied here:
these objects lie on the upper dashed portions of the curves in 
Fig.~\ref{fig5}. The collapse of very small objects with mass $<
M_{crit}$ is on the other hand made impossible by the 
cooling time being longer than the Hubble time (lower dashed
portion of the curves). Thus the only mass range in which 
negative feedback is important (solid portion) lies approximately 
in $10^6-10^8 M_\odot$, depending on redshift. In order for the negative
feedback to be effective, 
fluxes of the order of $10^{-24}-10^{-23}$erg s$^{-1}$ cm$^{-2}$
Hz$^{-1}$ sr$^{-1}$ are required. In turn, by using eq.~(\ref{si0}), we see that
these fluxes are produced by a Pop~III with baryonic mass $M_{b,5}$
at distances closer than $\simeq 21-7\times M_{b,5}^{1/2}$~kpc for the
two above flux values, respectively, while 
the SUVB can reach an intensity in the above range only after $z \approx 15$
(see \S~\ref{suvb}).
This suggests that at high $z$ negative feedback is driven primarily by the direct 
irradiation from neighbor objects in regions of intense clustering,
while only for $z \simlt 15$ the SUVB becomes dominant.

\subsection{Maximum incident flux}
\label{inc}

We define the maximum dissociating flux felt by a forming Pop~III as 
the sum $J_{LW,max}=(J_{LW,b}+J_{LW,d})$, 
where $J_{LW,b}$
is the SUVB and $J_{LW,d}$ is the direct flux produced by a nearby
luminous
object. $J_{LW,d}$ is defined as:
\begin{equation}
J_{LW,d}=\frac{\beta j_0}{4 \pi r^2};
\label{geid}
\end{equation}
$r$ is the distance between the forming object and the source of the dissociating
flux. The value of $j_0$ is the one given in eq. (\ref{gei0})
and the value of $\beta=1.5$ is adopted for the reasons explained in
\S \ref{star}.

We now derive the intensity of the SUVB. As the LW range
is very narrow ($\approx 2$~eV) we consider the average flux at the
central frequency of the band, $h \nu_0=12.45$ eV. To properly calculate the
intensity of the SUVB we must consider the intergalactic
H$_{2}$ attenuation including the effects of cosmological expansion and treat
in detail the radiative transfer through
LW H$_{2}$ lines.  LW lines are optically thin, 
with $\tau_{i}=N_{H_{2},i}\sigma_{i}\approx 0.05$, essentially for all the
lines. This
implies that as $J_{LW}$ is redshifted due to cosmological
expansion through LW lines, it is
attenuated by each line by a factor $e^{-\tau_{i}}$.
Abgrall \& Roueff (1989) have included in their study of
classic H$_2$ PDRs more than 1000 LW lines. As in our case H$_2$
formation, which leaves the molecule in excited roto/vibrational levels,
is negligible, a smaller number of lines $\approx 70$ -- involving the
ground state only -- needs to be considered.  Globally, $J_{LW}$ is
attenuated by a factor $e^{-\tau_{H2}}$,
where $\tau_{H2}=\sum_{i} \tau_{i}$, and $i$ runs up to the 71 lines
considered from the ground roto/vibrational states; we obtain $\tau_
{H_{2}} \simlt 3$, depending on the number
of lines encountered, and thus on the photon energy.
The intensity of the background is:
\begin{equation}
J_{LW,b}(z)=c \int^{z}_{z_{on}} \epsilon(\nu',z') \; {\cal A}_{H,H_2}
\frac{(1+z)^{3}}{(1+z')^{3}} \left\vert
\frac{dt}{dz'}\right\vert dz', 
\label{back}
\end{equation}
\begin{equation}
\label{epsi}
\epsilon(\nu',z')=\int^{M_{max}}_{M_{min}} j(\nu') {\cal N}(M,z') dM,
\end{equation}
where $z_{on}=29.6$ corresponds to the redshift at which the first 
objects start to form, $\nu'=\nu_0 (1+z')/(1+z)$; $\epsilon(\nu',z')$ [erg
cm$^{-3}$ s$^{-1}$ Hz$^{-1}$] is the proper emissivity due to sources
with $M_{min} \le M \le M_{max}$; ${\cal A}_{H,H_2}$ takes into account
the H and H$_2$ line absorption in the LW band and the H absorption
above the Lyman limit; the corresponding curves of growth are taken
from Federman, Glassgold \& Kwan (1979). 
$j(\nu')$ is taken from the BC spectrum as described in \S~\ref{star} with 
intensity given by eq.~(\ref{gei0}) 
and ${\cal N}(M,z')$ is the source number density in the mass interval
$dM$ derived from the PS formalism. 
The upper cut-off is $M_{max}=10^{14} M_{\odot}$, the
maximum statistically significant mass down to
$z=$8.3, the lowest redshift of the numerical simulation.
$M_{min}$ is the mass of the smallest halo that has been able to collapse
at the previous redshift step; we are thus assuming that all 
objects with masses greater than $M_{min}$ will contribute to the SUVB.
This calculation of $J_{LW,b}$ is correct as long as the filling factor of
the ionized and dissociated regions is small. As the volume filling factor
of these regions grows with time, their attenuation of the radiation
emitted by the sources is diminished and eventually, when the dissociation/ionization process is
complete, vanishes. To roughly take into account these complications,  
we assume that only a fraction $1-f_i$ of the IGM gas, where $f_i$ is the 
filling factor     of either ionized or dissociated regions, 
contributes to the attenuation.
We note that taking into account this effect, together with the determination
of the lower limit $M_{min}$ of the integral governed by the negative feedback, 
implies that eqs. (\ref{back})-(\ref{epsi}) cannot be solved   directly 
but must 
be computed iteratively as we derive the evolution of the reionization process. 

\section{STELLAR FEEDBACK}                      
\label{stfeed}

Once the first stars
have formed in the host protogalaxy, they can deeply influence the
subsequent star formation process through the effects of mass and energy deposition
due to winds and supernova explosions.
While low mass objects may experience a blowaway, expelling their entire gas content
into the IGM and  quenching star formation, larger objects may instead be able 
to at least partially retain their baryons.
However, even in this case the blowout induces a decrease of the star 
formation rate due to the global heating and loss of the galactic ISM. 
These two regimes are separated by a critical mass, $M_{by}$, to be
calculated in the following.    

For the relatively small objects present 
during the reionization epoch $30\simgt z \simgt 10$ the importance of these
stellar feedbacks can hardly be overlooked. To understand the role of stellar feedback, 
let us consider a collapsed object with mass lower than $M_{by}$.
Then the star formation is suddenly halted as the
entire gas content is removed. In this
case, the ionizing photon production will last only for a time interval of order
$t_{OB} \approx 10^7$ yr, the mean lifetime of the massive stars produced initially; 
after this, the ionized gas around the source will start to recombine at a
fast rate as a result of the highly efficient high $z$ Compton cooling, 
rapidly decreasing the temperature inside the ionized region. Due to the
short lifetime and  recombination time scales, these object will only 
produce transient HII regions which will rapidly disappear.
In fact, the recombination time scale, when the Compton cooling
is taken into account, is of order $t_{rec} \approx (1-50) \times 10^6$~yr 
at redshift $z$$\approx$30-10, respectively, 
and therefore much shorter than the corresponding Hubble time.
The dissociating photon production will nevertheless continue 
for a longer time, due to the important contribution of long-lived
intermediate mass stars formed in the same initial star
formation burst. This, combined with the fact that there is no
efficient mechanism available to re-form  the destroyed  H$_2$ in the IGM
analogous to H recombination, implies that the dissociation is not
impeded by blowaway and the contribution from these small 
objects should be yet accounted for.                   
However, this is not necessary. In fact, after blowaway, 
H$_2$ is efficiently formed in the
shocked IGM gas, cooling under non equilibrium conditions (Ferrara 1998). 
The final radius of the cooled shell behind which H$_2$ is formed
by this process is:
\begin{equation}
R_s \simeq 224  M_{b,5}^{1/5} (1+z)^{-19/10} \; {\rm kpc}.
\label{rs}
\end{equation}
We notice that the formation process produces an amount of
\HH roughly similar to the one destroyed by radiation (Ferrara 1998).
As inside 
$R_s$  \HH is re-formed, the net effect on the destruction of
intergalactic \HH of blown away objects is negligible (if not positive) 
and we can safely neglect them in the subsequent calculations. 

In conclusion, low mass objects with mass $< M_{by}$ 
produce ionization regions which last only for a recombination time;
larger objects can survive but their
star formation ability is impaired by ISM loss/heating. The rest of the Section 
is devoted to quantify these two cases.

\subsection{Low mass objects: $M < M_{by}$}
\label{lmo}

Due to the small mass of these proto-galaxies, 
massive star formation and, particularly, multi-supernova explosions
result in a blowaway. The blowaway is always preceded (unless the
galaxy is perfectly spherical) by a blowout, as explained in FT (see also MF).
The blowout and blowaway conditions can be derived by comparing
the blowout velocity, $v_b$, and the escape velocity of the
galaxy, $v_e$. 
From an analysis of the Kompaneets (1957) solution for the propagation
of a shock produced by an explosion in a stratified medium, it follows
that the shock velocity decreases down to a minimum occurring at about 
$3H$, where $H$ is the characteristic scale
height, before being reaccelerated; $v_b$ corresponds to such a
minimum.

To calculate these two velocities we  define a protogalaxy
as a two component system made of a dark matter halo and a gaseous disk.
We assume a modified isothermal halo density profile
$\rho_{h}(r)=\rho_c/[1+(r/r_0)^2]$
extending out to a radius $r_h = [3M_{h}/(4\pi (18 \pi^2) \rho_{crit})]^{1/3}$,
defined as the characteristic radius within which the mean
dark matter density is $18 \pi^2$ times the critical density 
$\rho_{crit}= 3 H_0^2 \Omega_0 (1+z)^3/8\pi G=1.88 \tento{-29} h^{-2} \,
(1+z)^3$ g cm$^{-3}$; $M_{h}$ is the halo mass.
For such a halo the escape velocity can be written as:
\begin{eqnarray}
\label{ve}
v_e &= &(2 \vert \Phi(r_d) \vert)^{1/2} \simeq ( 4\pi p G \rho_c 
r_0^2)^{1/2}  = \left({2 p G M_{h} \over r_h} \right)^{1/2}, \nonumber \\
    & \simeq & 0.9 M_{b,5}^{1/3} (1+z)^{1/2} 
    {\rm km~s}^{-1},
\end{eqnarray}
where $\Phi(r_d)$ is the the gravitational potential of the halo
(in this calculation we neglect the contribution of baryons to the 
potential)
calculated at the radius of the disk, $r_d$; 
$p$=1.65; the assumption $r_d \gg r_0$ has been made.

The explicit expression for $v_b$ has been obtained by FT:
\begin{equation}
\label{vb}
v_b = 2.7 \left({L\over H^2 \langle \rho_d \rangle}\right)^{1/3}~~{\rm
~km~s}^{-1},
\end{equation}
where $L$ is the mechanical luminosity and $\langle \rho_d \rangle$ is the disk mean gas
density. $L$ at redshift $z$  can be written as:
\begin{equation}
\label{L}
L={\epsilon_0\nu M_\star\over t_{ff}}\simeq
9.4\times 10^{35}(1+z)^{3/2} M_{b,5} f_{b,8} f_{\star,15}
{\rm ergs~s}^{-1},
\end{equation}
where $\epsilon_0=10^{51}$~ergs is the energy of a SN explosion;
as previously stated, we assume a Salpeter IMF, according to which one supernova
is produced for each 135~$M_\odot=\nu^{-1}$ of stars formed;
the free-fall time is $t_{ff}= (4\pi G \langle\rho_h\rangle)^{-1/2}$;

Due to the dissipative nature of the gas, the baryons which should be
initially
distributed approximately as the dark matter, will lose pressure and
collapse in the gravitational field of the dark matter. 
If the halo is rotating, the gas will collapse in a centrifugally
supported disk and , assuming that all the
baryons bound to the dark matter halo are able to collapse in the disk,
$\langle \rho_d \rangle \simeq (r_h^3 \Omega_b/ r_d^2 H)\langle\rho_h\rangle$. 
The radius of the disk can be estimated by imposing that the specific angular
momentum of the disk, $j_d$, is equal to the halo one, $j_h$ (Mo, Mao \&
White 1998; Weil, Eke \& Efstathiou 1998):
\begin{equation}
\label{jd}
j_h = \sqrt{2} \lambda v_c r_h = 2 v_c \ell_d = j_d,
\end{equation}
where $v_c$ is the circular velocity,
$\ell_d$ is the disk scale length and $\lambda$ the standard halo spin
parameter. As shown by numerical simulations (Barnes \& Efstathiou 1987;
Steinmetz \& Bartelmann 1995) $\lambda$ is only very weakly dependent on
$M_{h}$ and on the density fluctuation spectrum; its distribution is
approximately
log-normal and peaks around $\lambda=0.04$. From eq.~(\ref{jd}) we then
obtain
$\ell_d = (\lambda/\sqrt{2}) r_h$. For an exponential disk, as
implicitly assumed in
deriving eq.~(\ref{jd}) above,  the optical radius (\ie the radius
encompassing 83~\%
of the total integrated light) is 3.2 $\ell_d$; then, the HI
radius in galaxies
is typically found to be $\approx 2$ times larger than the optical radius 
(Salpeter
\& Hoffman 1996). Thus, for the radius of the gaseous disk we take:
\begin{equation}
\label{rd}
r_d = 4.5 \lambda r_h.
\end{equation}
The scale height, $H$, is roughly given by:
\begin{equation}
\label{H}
H = {c_s^2 r_d^2\over G M_h}=69 \lambda^2 \left({c_s\over v_e}\right)^2
r_h,
\end{equation}
where $c_s$ is the effective
gas sound speed, including also a possible turbulent contribution;
we take $c_s \approx 1$~km~s$^{-1}$, corresponding to the minimum
temperature allowed by molecular cooling.
It is useful to note that  $H^2 \langle \rho_d \rangle= 0.24 c_s^2 \Omega_b/G$,
\ie is independent of mass and redshift. We point out that these estimates are in excellent
agreement with the properties of Pop III disks found in numerical simulations of Bromm, Coppi \& Larson
(1999).

In the following we derive the necessary condition for blowaway to
occur. Following blowout, the pressure inside the hot gas bubble 
drops suddenly due to the inward propagation of a rarefaction wave. The
lateral walls of the shell, moving along the galaxy major axis will
continue to expand unperturbed until they are overcome by the
rarefaction wave at $r_c$, corresponding to a time $t_c$ elapsed from
the blowout. After that moment, the shell enters the momentum-conserving
phase, since the driving pressure has been
dissipated by the blowout. The requirement for the blowaway to take
place is then that the momentum of the shell (of mass $M_c$ at $r_c$)
is larger than the momentum necessary to accelerate the gas
outside $r_c$ at a velocity larger than the
escape velocity:
\begin{equation}
M_c v_{r_c} \geq M_h v_e.
\end{equation}
With the         assumptions and algebra outlined in FT, one can write
the condition for blowaway to occur:
\begin{equation}
\frac{v_b}{v_e} \geq (\epsilon-a)^2 a^{-2} {\rm e}^{3/2},
\label{baway}
\end{equation}
where $v_b$ is the blowout velocity, $v_e$ is the escape velocity, $a$
is a parameter equal to 2/3 and $\epsilon=r_d/H$ is the ratio
of the major to the minor axis of the proto-galaxy. From eqs.~(\ref{rd})
and~(\ref{H}), we find:
\begin{equation}
\epsilon^{-1}= 15.3 \lambda \left( \frac{c_s}{v_e} \right)^2.
\label{eps}
\end{equation}
Substituting eqs.~(\ref{ve}),~(\ref{vb}) and~(\ref{eps}) into
eq.~(\ref{baway}), we find the critical      mass below which blowaway occurs. 
It is useful to recall that this prescription has been carefully tested against hydrodynamical
simulations presented in MF.

In Fig.~\ref{fig6} $M_{by}$ and
$M_{crit}$ are plotted for the usual cosmological parameters
($\Omega_0$=1, $h$=0.5 and $\Omega_b$=0.06) and for the three runs
A, B, C, whose parameters are defined in Tab. 1 and described in
detail in \S \ref{res}. As $M_{by}$ depends on the product 
$f_b \times f_\star$ all other runs give the same results as run A.
As it is seen from the Figure, the blowaway mass depends very weakly 
on this product (approximately $\propto (f_bf_\star)^{1/4}$),
and it is always larger than $M_{crit}$, but
the number of objects suffering blowaway decreases with redshift.
Note as that the vast majority of the objects which virialized at redshifts
$\simgt 15$ will be blown away due to their low mass predicted by 
hierarchical models.  
We also point out that at redshifts below $\approx$6, when objects
with masses $\ge 10^9 M_\odot$ start to dominate the mass function, 
our estimate of $M_{by}$ becomes less accurate. This is due to 
the assumption implicitly made that all SNe drive the same superbubble.
Although valid for small objects, 
for larger galaxies this idealization becomes increasingly poorer and one should        
consider the fact that in general SNe are distributed among 
OB associations with different values of $L$ spread around the galactic
disk. However, our calculations are not affected by this problem 
as we stop the simulations at $z \approx 8$.

\subsection{Larger objects: $M > M_{by}$}
\label{lo}

Objects with $M>M_{by}$ will not experience blowaway and rather than explode,
they continue to produce stars quiescently.  Nevertheless,
a blowout with consequent mass loss and heating of the ISM by 
SN shock waves will very likely take place.  
MF found  that galaxies with masses
lower than a few $\times 10^{10}$M$_\odot$ will suffer losses in an
outflow, although the gas loss efficiency is not very high. Most important,
perhaps, is the evaporation and destruction of molecular clouds as a 
result of an enhanced star formation activity which depresses the    
star formation rate. These processes are not yet fully understood in their
complexity and detailed hydrodynamical models implementing the physics
regulating the multi-phase structure of the ISM are required to make further
progress. We parameterise these processes by 
an heuristic approach prescribing that the mass transformed into   
stars is lower than  the value $M_\star$ used for low mass objects 
by a mass dependent factor $s_{feed}$:
\begin{equation}
M_\star^\ell=s_{feed}M_\star=\frac{1}{1+(M_{by}/M)^\gamma}M_\star ,
\label{sf}
\end{equation}
where $\gamma$ is a free parameter usually taken equal to 2. 
We have checked that our results are almost insensitive to different values of $\gamma$.
This parameterization of the stellar feedback is analogous to 
the one typically used in semi-analytical models of galaxy formation
(White \& Frenk 1991; Kauffman 1995; Baugh \etal 1998a; Guiderdoni \etal 1998).
This factor allows for a maximum 50\% decrease of the stellar mass of a
galaxy due to gas heating/loss; this stellar feedback becomes increasingly 
less important for larger objects. 

\section{Summary of Evolutionary Tracks}
\label{stracks}
Given the large number of processes so far discussed and included in the model, 
it is probably worth to summarize them briefly before presenting the results.
Fig. \ref{fig7} illustrates all possible evolutionary tracks and final fates     
of primordial objects, together with the mass scales determined by the various 
physical processes and feedbacks. We recall that there are four critical mass
scales in the problem: (i) $M_{crit}$, the minimum mass for an object to be 
able to cool in a Hubble time; (ii) $M_H$, the critical mass for which hydrogen
Ly$\alpha$ line cooling is dominant;
(iii) $M_{sh}$, the characteristic mass 
above which the object is self-shielded, and (iv) 
$M_{by}$ the characteristic mass for stellar feedback, below which blowaway 
can not be 
avoided. Starting from a virialized dark matter halo, condition (i) produces 
the first branching,  and objects failing to satisfy it will not collapse
and form only a negligible amount of stars. In the following, 
we will refer to these objects as {\it dark objects}. 
Protogalaxies with masses in the range $M_{crit} <
M < M_{H}$ are then subject to the effect of radiative feedback, 
which could either impede the collapse of those of them with 
mass $M<M_{sh}$, thus contributing to the 
class of dark objects, or allow the collapse of the remaining ones ($M>M_{sh}$)
to join those with $M>M_H$ in the class of {\it luminous objects}. This is the 
class of objects that convert a considerable fraction of their baryons in stars.
Stellar feedback causes the final bifurcation by inducing a blowaway of 
the baryons
contained in luminous objects with mass $M<M_{by}$; this separates the class in
two subclasses, namely "normal" galaxies (although of masses comparable to 
present day
dwarfs) that we dub {\it gaseous galaxies} and tiny stellar aggregates with
negligible traces (if any) of gas to which consequently we will refer to as
{\it naked stellar clusters}. The role of these distinct populations for the
reionization of the universe and their density evolution will be clarified
by the following results.

\section{RESULTS}
\label{res}

In this Section we present the results obtained by including all the
effects discussed above in the calculation of the reionization history.
As the cosmological model (CDM with $\Omega_0$=1, $h$=0.5, $\Omega_b$=0.06)
is fixed by the N-body simulations, there are three free parameters left,
namely $f_b$, $f_{\star}$ and $f_{esc}$, defined in \S~\ref{star}
(actually, as already pointed out, the number of effective free parameters
is two, $f_b\times f_{\star}$ and $f_{esc}$).
We have performed runs with different values for these parameters, labelled as runs
A-D in Table~\ref{tab1}; our reference case is run A ($f_b=0.08$,
$f_\star=0.15$ and $f_{esc}=0.2$). We have made runs
with the same parameters as in run A, but in which only
objects with masses $M>M_H$ are considered (run A1) and the normalization
of the fluctuation spectrum has been varied (run A2).
These cases are intended to isolate     the effects of negative feedback
and of a different evolution of the dark matter halos, as explained below. 

The present approach, based on a combination of very high resolution 
N-body numerical simulations and detailed treatment of the radiative/stellar
feedbacks, allows us to study the inhomogeneous ionization process and
the nature of the sources producing it at a remarkable detail level. 
The main output of our model consist of simulation boxes
in which ionized/dissociated regions can be clearly identified and for 
which we can derive the global three dimensional structure 
as a function of cosmic time.

\subsection{Filling factor}
\label{ff}

As a typical example, we show in Fig.~\ref{fig8} the actual outcome for run
A. There, the  spatial distribution of 
reionized regions is shown at four different redshifts ($z=22.1, 19.8, 
18.0, 15.4$). The spheres grow in number and volume with time,     although
from $z\approx$15 the number of luminous objects flattens. The reason 
for this behavior is that the population is dominated by 
small mass objects whose collapse is prevented either by the condition 
$M>M_{crit}$ or by the increased intensity of the SUVB 
(see \S~\ref{gtracks}). The images shown in
Fig.~\ref{fig8} give an immediate and qualitative view of the evolution
of the volume occupied by ionized atomic hydrogen. A more quantitative
measure of the filling factor of the ionized gas can easily be obtained.

The filling factors of the dissociated H$_2$ and ionized H are defined as 
the box volume fraction occupied by those species. 
The results are shown in Fig.~\ref{fig9}a and~\ref{fig9}b, for different
runs. The intergalactic relic molecular hydrogen is found to be  
completely dissociated at very high redshift ($z\approx 25$) independently
of the parameters of the simulation. This descends from the fact that 
dissociation spheres are relatively large and overlap at early times.
Ionization spheres are instead always smaller than dissociation ones 
and complete reionization occurs considerably later. 
Except for run C, when reionization occurs by $z$$\approx$15,
primordial galaxies are able to reionize the IGM at a redshift
$z$$\approx$10. 
The filling factor is approximately linearly 
proportional to the number of sources but has a stronger (cubic) dependence
on the ionization sphere radius. Thus, although the number 
of (relatively small) luminous objects present increases with redshift, 
the filling factor is only boosted by the appearance of larger, and therefore
more luminous, objects which can ionize more efficiently. 
The subsequent flattening at lower redshifts, is obviously due to the 
fact that when the volume fraction occupied by the ionized gas becomes
close to unity most photons are preferentially used to sustain the reached 
ionization level rather than to create new ionized regions.

In principle, a higher photon injection in the IGM could result both in an
increase (as larger HII regions are produced) or a
decrease (as the number of sources is reduced by the effect of radiative
feedback) of the filling factor. Along the sequence run B, D, A, C the
number of ionizing photons injected in the IGM is progressively increased.
Then Fig.~\ref{fig9}a allows us to conclude that the former effect is
dominating, \ie the radiative feedback is of minor importance.

For the specific case of run A (we do not discuss in further detail He
reionization in this paper)
we show the analogous filling factor for the doubly ionized He;
the lowest curve in Fig.~\ref{fig9}a shows its evolution.
As $R_{s,He}\simeq0.1 R_{s}$ (see eqs.~[\ref{rion}] and~[\ref{rihe}]), the
helium filling factor should be approximately 0.1 \% of the hydrogen
filling factor, but it increases when the effects of the
ionized hydrogen sphere overlapping is included (see Fig.~\ref{fig9}a).
The stellar spectrum contributed by the reionizing galaxies 
is relatively soft and the number of photons above the He$^{++}$ 
ionization threshold very limited. Hence, He reionization does not 
occur in our model and could be possibly achieved only through 
quasar-like sources with harder spectra.

In Fig.~\ref{fig9}b we analyze the effect of negative feedback and the
normalization of the dark matter fluctuation power spectrum,
$\sigma_8$. By comparing run A with run A1, in which only
objects with masses $M>M_H$ are considered,  we see that, apart from
very high redshifts where few such objects are present and the bulk of
photon production is due to objects with $M<M_H$, the main contribution
to the IGM reionization comes from sources with $M>M_H$ and Pop~IIIs (with
$M<M_H$) alone would not be able the reionize the IGM. This  statement
remains valid even in the absence of radiative feedback effects on Pop~III
as concluded from the results of a run A in which radiative feedback has not 
been allowed. In this case, although a slightly higher 
filling factor is obtained as the contribution from previously 
suppressed Pop~III is now present,  
the difference with run A is never higher than 10\%. 

Finally, in run A2, the normalization of the fluctuation spectrum 
has been decreased to $\sigma_8=0.5$. The evolution of the
filling factor appears similar   to that in run A, 
but shifted towards lower redshift. This is due to the delayed evolution
of this model with respect to the reference one, which results in a lower
number of luminous objects at a given redshift. 

\subsection{SUVB}
\label{suvb}

The derived evolution of the SUVB is presented in 
Fig.~\ref{fig10}a at the mean frequency of the LW band,
$h \nu_0$=12.45 eV, for runs C, A, D and B (from the top to the bottom). 
The intensity scales with the product $f_b \times f_\star \times f_{esc}$ 
as deduced from eqs.~(\ref{gei0}),~(\ref{back}) and~(\ref{epsi}).
Typically, a flux lower than 10$^{-24}$ erg cm$^{-2}$ s$^{-1}$ Hz$^{-1}$
sr$^{-1}$ is found at $z$$\simgt$15, except for case C for which it
is found to be about one order of magnitude larger. As an intensity of at least few
$\times 10^{-24}$ erg cm$^{-2}$ s$^{-1}$ Hz$^{-1}$ sr$^{-1}$ at the Lyman limit
is needed for an appreciable negative feedback at these redshifts
(see Fig.~\ref{fig5}), we conclude, in agreement with CFA, that 
the SUVB flux is too weak to affect small mass structure formation
at high redshift. At these early epochs, negative feedback is
instead induced by the direct flux from pre-existing nearby
objects. The SUVB becomes intense and dominates the direct flux 
for $z$$<$15 (earlier for run C) 
and consequently it governs the formation of late forming Pop~IIIs.
The additional cases for runs A1 and A2 are shown in    
Fig.~\ref{fig10}b.  Run A1 produces the same SUVB as run
A at low redshift; at early times, where there is little contribution from 
objects with $M>M_H$, the SUVB intensity is lower.  
The case with a lower normalization (run A2) also produces a lower SUVB, as expected.

\subsection{SFR}
\label{SFR}

We have calculated the star formation
rate per comoving volume (SFR) predicted by our model and we have
compared it with the values derived by the
most recent studies at lower redshift (Lilly \etal 1996;
Connolly \etal 1997; Madau, Pozzetti \& Dickinson 1998; 
Steidel \etal 1998). Inside each              simulation box we
have derived the SFR as $SFR=V^{-1} \sum_i sfr_i$, where $V$ is the
comoving volume of the simulation box and $sfr_i$ are the star formation
rates of the individual luminous objects:
\begin{equation}
sfr_i= \frac{M_{\star,i}}{t_e},
\label{sfr}
\end{equation}
in units of M$_{\sun}$ yr$^{-1}$. Here $M_{\star,i}$ is given by
eq.~(\ref{mstar}), showing that the SFR crucially depends on the parameters
$f_b$ and $f_{\star}$; $t_e$ is defined as the free-fall time
for objects with mass $M>M_{by}$, while for small mass objects
($M<M_{by}$), witnessing only a single burst of star formation,
we assume it equal to the Hubble time:
\begin{equation}
t_e= \left\{
\begin{array}{ll}
t_H & M < M_{by},\\
t_{ff} & M \ge M_{by},\\
\end{array}\right.
\label{te}
\end{equation}
where $t_H=(2/3) H_0^{-1} \, (1+z)^{-3/2}$.
As we have already pointed out, the simulations, although resolving small
mass objects, might miss halos larger than $M_{max,s}$; this value 
increases with decreasing redshift. To take into account this limitation
and correct for it, we include also
objects with masses $M>M_{max,s}$ and model their contribution 
as follows. At each redshift, this contribution is given by:
\begin{equation}
sfr_{PS}=\int_{M_{max,s}}^{M_{max}} \frac{M_\star(M)}{t_e(M)} {\cal N}(M)
dM,
\label{sfrps}
\end{equation}
where $M_\star$ is given in eq.~(\ref{mstar}), while ${\cal N}(M)$ and
$M_{max}$ are defined in eq.~(\ref{epsi}). We implicitly assume
that these objects do not suffer from radiative feedback (this assumption is
justified by the fact that these are usually objects with masses
$M>M_H$) and that these halos host luminous objects.     
We then add this contribution to the SFR derived from the simulations.
The results are shown in
Fig.~\ref{fig11}a for runs A-D.
Obviously, the higher the $f_b \times f_\star$
product, the higher SFR is obtained. Actually, although run A and D
use  the same values for the above parameters, in run D more luminous
objects are formed (as the SUVB is lower [see \S~\ref{suvb}] more
objects escape negative feedback) and this results in a slightly higher
SFR. We compare the curves obtained with the most recent measurements of
the SFR, namely: Lilly \etal (1996) [circles], Madau, Pozzetti \&
Dickinson (1998) [triangles], Connolly \etal (1997) [squares] and
Steidel \etal (1998) [crosses]. The recent lower limit to the SFR at
$z$$\approx$3 
set by       SCUBA (Hughes \etal 1998) coincides with the point
corresponding to Steidel \etal (1998).
The points from Madau, Pozzetti \& Dickinson (1998) are
derived from a study of the Lyman break galaxies  in the Hubble Deep
Field (HDF) and show a decline of the SFR at high redshift. The 
measurements by Steidel \etal (1998) do not
show the same decline and favour a more or less constant SFR at high
redshift. This determinations are based on a study 
of star-forming galaxies at $z>$3.8 in a field much wider than the HDF,
although shallower. All points are corrected for dust extinction as in
Steidel \etal (1998).
From Fig.~\ref{fig11}a, we see that runs B and C can be taken
respectively as  lower and upper limits to the SFR obtained from our
model, thus constraining the product $f_b \times f_\star$.
Run A and D produce a trend which apparently well matches
the observations, suggesting a likely    value of        the product $f_b \times
f_{\star}$ around $\approx$0.01. In Fig~\ref{fig11}b we study the effect
of negative feedback and $\sigma_8$. From a comparison between run A and
A1, where only objects with $M>M_H$ are considered, we see that, while
at high redshift the main contribution to the SFR comes from small
mass objects ($M<M_H$), at lower redshift, their contribution becomes
negligible. Finally, run A2, with $\sigma_8$=0.5, produces a lower SFR
as less luminous objects are present at a given redshift, again the same
point made in \S~\ref{ff}.

Additionally, we have derived the evolution of the stellar density
parameter,
$\Omega_\star(z)=\rho_{\star,tot}(z)/\rho_{crit}$, produced by run A;
$\rho_{\star,tot}(z)$ is the
total mass density in stars at redshift $z$ inside the comoving volume of the
box $V$=125 Mpc$^3$, and is given by
\begin{equation}
\rho_{\star,tot}(t)= \int_{t_{on}}^{t} SFR(t) dt,
\end{equation}
where $t_{on}$ is the Hubble time corresponding to $z_{on}$=29.6.
We find that by the time the IGM is completely
reionized, $z$$\approx$10, only 2\% of the stars observed at $z=0$ 
($\Omega_\star(0)\approx4.9 \times 10^{-3}$ [Fukugita, Hogan \& Peebles
1998]) has been
produced. This apparently small amount of stars is sufficient to produce
the necessary ionizing flux. The number of ionizing photons
produced by $z$$\approx$10 can be derived as follows.
At $z$$\approx$10, $M_{\star,tot}\approx 3.8 \times 10^8$ M$_\odot$,
corresponding to an ionizing photon rate $S_i(0)\approx 2.62 \times
10^{54}$ s$^{-1}$ for the standard case A (see eq.~[\ref{si0}]).
The time at which the
lowest mass OB stars expire is $\approx$40 Myr (Oey \& Clarke 1997), thus, the
total number of ionizing photons escaping from luminous objects
is N$_i \approx 3.3 \times 10^{69}$. As the number of
baryons in the box is N$_b=V \Omega_b \rho_{crit}/(\mu m_p) \approx 4.7
\times 10^{68}$, where $\mu\approx 1.27$ is the mean molecular weight, there are
N$_i$/N$_b$$\approx$7 ionizing photons per baryon available.
At $z$$\approx$10 the
Hubble time is $t_H\approx 3.59 \times 10^8$ yr, while the mean hydrogen
recombination time over the time interval $z$$\approx$30-10 of interest
here is $\langle t_{rec} \rangle \approx 1.33 \times 10^8$ yr. Thus,
typically an atom has recombined about $t_H/\langle t_{rec} \rangle \approx$3
times. However, the ionizing photon budget provided by the formed stars is
sufficient to balance this recombination and keep the gas ionized.

A rough estimate of the mean metallicity produced by
these stars can be calculated. Given the adopted IMF and the total mass in stars at
$z$$\approx$10, $M_{\star,tot}$, we can derive the number of SNe in the
simulation box, and, assuming that each SN produces $\approx$1 $M_{\odot}$ of heavy
elements, we find that the total mass of heavy elements at $z$$\approx$10 is
$M_{met}\approx 2.81 \times 10^6 M_\odot$. As the total mass in baryons in the
considered cosmic volume is
$M_{b,tot} \approx 5.06 \times 10^{11} M_\odot$, we find that the mean
metallicity at $z$$\approx$10 is:
\begin{equation}
\langle Z\rangle =\frac{M_{met}}{M_{b,tot}}\approx 5.5 \times 10^{-6} 
\approx 3 \times 10^{-4} Z_\odot.
\label{met}
\end{equation}

\subsection{Galaxy evolutionary tracks}
\label{gtracks}

In this Section we discuss 
the final fates of primordial objects and we show their relative numbers in Fig.~\ref{fig12}.
The straight lines represent, from the top to the bottom,
the number of dark matter halos, dark objects, naked
stellar clusters and gaseous galaxies, respectively.
The dotted curve represents the
number of luminous objects with large enough mass
($M>M_H$) to make the H line cooling efficient and become insensitive to
the negative feedback. We remind that the naked stellar
clusters are the luminous objects with $M<M_{by}$, while the
gaseous galaxies are the ones with $M>M_{by}$; thus, the number of
luminous objects present at a certain redshift is given by the sum of
naked stellar clusters and gaseous galaxies.
We first notice that the majority of
the luminous objects that are able to form at high redshift will
experience blowaway, becoming naked stellar clusters, while only a minor
fraction, and only at $z\simlt$15, when larger objects start to form,
will survive and become gaseous galaxies.
An always increasing number of luminous objects is forming with
decreasing redshift, until $z$$\approx$15, where a flattening is seen. This
is  due to the fact that the dark matter halo mass function is
still dominated by small mass objects, but a large fraction of them cannot
form due to the following combined effects: i) towards lower
redshift
the critical mass for the collapse ($M_{crit}$) increases and fewer
objects satisfy the condition $M>M_{crit}$; ii) the radiative feedback
due to either the direct dissociating flux or the SUVB (see \S~\ref{rf})
increases at low redshift as the SUVB intensity reaches values
significant for the negative feedback effect.
When the number of luminous objects becomes 
dominated by objects with $M>M_H$, by
$z$$\approx$10 the population of luminous objects grows again, basically because their
formation is now unaffected by negative feedback. 
A steadily increasing number   of objects is prevented from forming stars
and remains dark; this population is about $\approx$99\% of the total
population of dark matter halos at $z$$\approx$8. This is also due to the
combined effect of points i) and ii) mentioned above.
This population of halos which have failed to produce
stars could be identified with the low mass
tail distribution of the dark galaxies that reveal
their presence through gravitational lensing of quasars (Hawkins 1997;
Jimenez \etal 1997). It has been argued that this
population of dark galaxies outnumbers normal galaxies by a substantial
amount, and                 Fig.~\ref{fig12} supports this view.
At the same time, CDM models predict that 
a large number of satellites (a factor of 100 more than
observed [Klypin \etal 1999; Moore \etal 1999b]) should be present around normal galaxies.  
Many of them would 
form at redshifts higher than five and would survive merging and tidal
stripping inside larger halos to the present time. Their existence is
then linked to that of small mass primordial objects and the natural
question arises if these objects can be
reconciled with the internal properties of halos of present day-galaxies.

A question that naturally arises is which is the fate of the naked
stellar objects. The actual number of naked stellar objects should not
be much higher than the one at $z\approx$8, for two reasons: (i) with
decreasing redshift, a lower number of luminous objects is subject to blowaway,
as increasingly larger objects form; (ii) as already pointed out, at
redshifts below $\approx$6, our estimate of $M_{by}$ becomes less
accurate, actually larger than the real value (see FT), thus, the number
of luminous objects undergoing a blowaway is decreased also by this
effect. With this caveat, approximately 2 naked stellar objects should end up
in the Galactic halo, where, at present time, only stars with masses in
the range 0.1-1 $M_\odot$  are still shining. As the maximum total mass of a
naked stellar object at $z\approx$8 is $\approx 10^9$ $M_\odot$, given a
Salpeter IMF, we find that the upper limit for the number of the above
relic stars is $\approx 4 \times 10^6$. As in the Galaxy, given the same
IMF, $\approx 2 \times 10^{11}$ stars with masses in the range 0.1-1
$M_\odot$ are present, one out of $\approx 5 \times 10^4$ stars
comes from naked stellar objects. The upper limit on the metallicity of such
stars is the IGM one at $z\approx$10, \ie $Z\approx 3 \times 10^{-4} Z_\odot$,
as derived in eq.~(\ref{met}).

We have compared the results for runs A and C.
The only significant difference is that
in case C, where a higher SUVB is found,   the
redshift at which the $M>M_H$ population peaks               is higher,
as from $z  
\approx 15$ there are very few objects that can avoid negative feedback
and the only surviving halos are the ones for which H line cooling is efficient.

\section{Comparison with previous work}

In the last decade there have
been a number of different works dedicated to reionization. The ``first generation''
of studies used very simplified descriptions of the problem in terms of the treatment of the
physical processes, cosmological evolution of the ionizing sources, or both. 
These pioneering works (Fukugita \& Kawasaki 1994; Tegmark, Silk \& Blanchard 1994;
Giroux \& Shapiro 1996) found results often in disagreement among them, probably
due to the different effects/cosmological models included or neglected.  
More recent works have tackled again the problem improving on at least two 
crucial ingredients: a detailed cosmological hydrodynamic evolution of the 
IGM/galaxy
formation, which gives a more detailed distribution of ionizing sources and of the
gas, and the inclusion of newly discovered feedback effects, along the lines 
presented here. GO performed detailed simulations for
a CDM+$\Lambda$ cosmological model with an approximate treatment of the
radiative transfer and allowing for IGM clumping. 
The N-body simulations used here have a larger number 
of particles (256$^3$ instead of their 64/128$^3$) although the size of the
box is practically the same. Their spatial resolution ($\approx$1~kpc) can only
marginally resolve the HII spheres which, for Pop III objects, are often smaller than
that size. Also, we treat in more detail the effects of radiative and stellar
feedback, even if we cannot compete with the wealth of information on the IGM
density distribution that can be extracted from their simulations. They conclude that 
the reionization redshift is $\approx$7, \ie about 3 redshift units below the
one found here. The discrepancy might depend both on the different cosmological model
adopted and/or on the above differences. A low value for the reionization
redshift is found also in the semi-analytical model presented by
VS: for both an open and a critical density universe,
they use a mass function alternative to the usually adopted PS formalism and,
given a prescription for galaxy formation and photon production, they find that
the reionization is complete by redshift 6.8 for the open universe and 5.6 for
the critical density one. As in their calculation they include the treatment of
the gas clumpiness, this could again explain the discrepancy with the value
found in the present work. Additionally, deriving the radius of the HII region
surrounding a source, they neglect the influence of the expansion of the
universe, underestimating the real radius and obtaining a late
reionization epoch. 
Haiman \& Loeb (1997) used a semi-analytical model based on PS formalism 
which adopted a strong version of the radiative feedback (but not considering
the stellar one), assuming the suppression of star formation inside
objects with virial temperature below $10^4$~K. They also neglect the
effect of gas clumpiness. This approach has the advantage 
of being very flexible when exploring the dependence of reionization on various
parameters, but clearly lacks the crucial information about the spatial distribution
of ionizing sources. They find  that for a large range of CDM models
reionization occurs at a redshift $\simgt$10, consistent with the value we
find in the present work. The basic agreement with their result derives from
the fact that we also find that reionization is mostly driven by objects collapsed
through H line cooling. Stellar feedback instead introduces a different prediction
for the fate of ionizing objects as discussed above.

\section{DISCUSSION}
\label{disc}

In this paper we have studied the reionization of the universe due to an 
inhomogeneous distribution of sources, including 
a number of relevant physical processes and feedback effects whose 
importance for this type of studies has only recently been recognized.  
Our approach provides a reliable picture
of the actual process of IGM reionization.

Maybe, the major feature not considered here is constituted by the possible 
IGM density inhomogeneities, as throughout the paper we have assumed an
homogeneous gas distribution. 
The IGM clumpiness can have in principle different effects on the
reionization of the universe. Both GO and VS have studied
the process of reionization 
taking into account the effect of the inhomogeneous gas distribution through a
clumping factor $C=\langle \rho_{IGM}^2\rangle/\langle\rho_{IGM}\rangle^2$,
that, included in the calculation of the recombination
process, increases
the number of ionizing photons necessary to reionize the IGM 
by the amount $C$, thus delaying the IGM complete reionization 
with respect to the homogeneous case.
The gas clumpiness influences also the shape of
the ionized regions: indeed, the propagation of the ionizing front in an
inhomogeneous medium, does not result in a spherically symmetric HII
region, with ionization proceeding in an inside-out fashion, but rather
the reionization occurs outside-in, starting in voids and
gradually penetrating in high-density regions as pointed out by Miralda-Escud\'e, 
Haehnelt \& Rees (1998). If this is the case, recombinations from overdense
regions which are presumably very compact, would contribute to the average 
recombination rate only at later times, when most of the volume has already 
been ionized. Note that this effect is opposite to the first one,
indicating that at a given redshift in the homogeneous case one
 might be underestimating the filling
factor of the ionized gas.
An even more refined approach to the problem should include
the combined treatment of the complete radiative transfer equation in an
inhomogeneous medium. The very first steps in this directions have been presented
by Norman, Paschos \& Abel (1998), who are trying to incorporate radiative
transfer into 3D hydrodynamic cosmological simulations.
As a preliminary study of the effect of the gas clumpiness on our results, we have 
made additional runs in which we include a clumping factor in the
calculation of the recombination process. For $C$ we have adopted the curves
given in GO and VS.
In Fig.~\ref{fig13} the HII filling factor for run A is
shown, together with the curves derived  in the presence of 
clumping. The GO clumping factor (dashed
curve) produces little difference, as it is close to unity until
$z\approx$10. On the other hand, the VS one (dotted
line), reaching much higher values, results in a substantially lower
filling factor and a delay in the IGM reionization, which is not
complete by $z$$\approx$8.

Another important output of the model is the star formation history.
A direct comparison with most of the previous calculations of the SFR, 
both in numerical and semi-analytical approaches, can only be
approximate, as those mainly concentrated on 
lower redshift range than the ones
studied here          (see for example Guiderdoni \etal 1998; 
Baugh \etal 1998b; Pei, Fall \& Hauser 1998; Nagamine, Cen \& Ostriker 1999).
VS, on the other hand, calculate the SFR up to 
$z$$\approx$20, obtaining values slightly lower both than ours and the 
ones observationally derived.

Finally, the result that a high fraction of dark objects is present, is
intriguing, as the question of whether dark galaxies can exist is
a longstanding one (Dekel \& Silk 1986). Indeed, the fact that
the luminosity function has a flatter faint slope than the associated 
halo mass function can be explained with an increasing fraction of dark
galaxies towards lower masses. Some authors (Babul \& Rees 1992;
Metcalfe \etal 1997) claim that the initial burst of the star formation
in small mass objects may provide a population of
disappearing dwarfs proposed for interpreting the faint galaxy counts.
On the other hand, Jimenez \etal (1997) propose that a dark galaxy forms
as a low-density disk in a dark halo of high spin parameter: such a
galaxy may have a surface density too low to produce a significant
star formation rate. Our model offers an alternative explanation to 
the nature of such objects. 

A comparison of the predictions of our model is currently possible
only in a indirect way, as done throughout the paper for several 
quantities. However, in a near future, thanks to the availability 
of powerful spatial and ground-based instrumentation it will become
possible to directly probe most of the activity occurring prior to
complete reionization. For example, NGST is expected to produce the    
detection of a large number of high-z SNe (Marri \& Ferrara 1998) 
that could be used to 
trace early star formation and stellar feedback in Pop III objects, 
and possibly the IGM, via absorption experiments. 
NGST will be also 
able to determine with great accuracy the reionization epoch 
from the spectra of high-redshift sources (Haiman \& Loeb 1998c). 
The ionized regions can be mapped directly in free-free emission (Oh
1999) by the Square Kilometer Array (SKA).
Reionization should
leave a measurable imprint on the CMB, as secondary anisotropies
are produced via the kinetic Sunyaev-Zeldovich effect by inhomogeneous
reionization (Knox, Scoccimarro \& Dodelson 1998; Gruzinov \& Hu 1998); this experiment is well
into the capabilities of MAP and Planck missions.  
submillimetre telescopes will be used to search for dust, produced by
primordial objects, and possibly molecules; the dust can be an important
reservoir of the metals synthesized by the first SNe and it might
have important consequences for the exploration of the early
universe. Finally, radio surveys might reveal signatures of reionization
and give insight into the thermal evolution of neutral hydrogen
through its redshifted HI 21-cm line emission. These features are 
within reach of new generation radio telescope like SKA (Madau, Meiksin
\& Rees 1997; Tozzi \etal 1999).  Clearly, exciting times are ahead of us.

\section{SUMMARY}
\label{sum}

We have studied inhomogeneous  reionization in a critical density CDM
model, including a detailed
treatment of radiative and stellar feedback processes on galaxy  
formation.
The main results discussed in this paper can be summarized as 
follows.
\begin{itemize}
\item Galaxies are able to reionize the neutral atomic hydrogen by
a redshift $z$$\approx$10, while molecular hydrogen is completely
dissociated at very high redshift ($z$$\approx$25).
\item IGM reionization is basically driven by objects collapsed through
H line cooling ($M>M_H$), while small mass objects ($M<M_H$) play only a
minor role and even in the absence of a radiative negative feedback
they would not be able the reionize the IGM.
\item The soft-UV background intensity is too low to produce sensible negative 
feedback effects on low   mass galaxy    formation at redshift
$z$$\simgt$15, where the radiative feedback is dominated by the direct flux
from pre-existing objects. At lower redshifts the SUVB is instead
dominant and reaches interesting values to influence 
the subsequent galaxy formation process.
\item The evolution of the star formation rate obtained shows a trend
consistent with the most recent measurements and the match constrains 
the baryon conversion efficiency into stars
in a narrow range around $\approx$0.01. Only about 2\% of the stars 
observed at $z=0$ is required to reionize the universe. This corresponds to an
average IGM metallicity at redshift $z$$\approx$10 equal to $\langle Z
\rangle \approx 3 \times 10^{-4} Z_\odot$.
\item A consistent         fraction of halos   is prevented from forming stars
by either the condition $M<M_{crit}$ or the effect of radiative feedback;
this population of dark objects reaches $\approx$99 \% of the
dark matter halo population at $z$$\approx$8.
\end{itemize}
\acknowledgments{We thank Z. Haiman, L. Pozzetti and  M. Tegmark for providing
useful data; D. Galli, F. Haardt, F. Palla and  M. Rees for 
stimulating discussions. 
A special thank goes from BC to A. Riccardi for his help with software.
The work presented in this paper was carried out using data made available 
by the Virgo Supercomputing Consortium 
(http://star-www.dur.ac.uk/~frazerp/virgo/virgo.html) 
and   computers based at the Computing Centre of the Max-Planck
Society in Garching and Edinburgh Parallel Computing Center.}

\newpage
\begin{center}
\Large Tables 
\vskip 8.cm
\end{center}

\begin{table}[hp]
\centerline{\begin{tabular}[t] {|l|l|l|l|l|l|l|r|} \hline
          &             &                   &                 &
	 &             \\        
{\em RUN} & {\em $f_b$} & {\em $f_{\star}$} & {\em $f_{esc}$} &
{\em cooling} & {\em $\sigma_8$}\\
          &             &                   &                 &
	 &              \\ \hline 
A  & 0.08 & 0.15 & 0.2 & H+H$_2$ & 0.6\\ 
B  & 0.08 & 0.05 & 0.2 & H+H$_2$ & 0.6\\
C  & 1.00 & 0.15 & 0.2 & H+H$_2$ & 0.6\\
D  & 0.08 & 0.15 & 0.1 & H+H$_2$ & 0.6\\
A1 & 0.08 & 0.15 & 0.2 & H       & 0.6\\
A2 & 0.08 & 0.15 & 0.2 & H+H$_2$ & 0.5\\ \hline
\end{tabular}}
\caption{Parameters of the calculation: fraction of virialized baryons
that are able to cool and be available for star formation, $f_b$; star
formation efficiency, $f_{\star}$; photon escape fraction, $f_{esc}$;
object cooling mechanism; dark matter
fluctuation power spectrum normalization, $\sigma_8$.}
\label{tab1}
\end{table}

\newpage

\begin{figure}[t]
\centerline{\psfig{figure=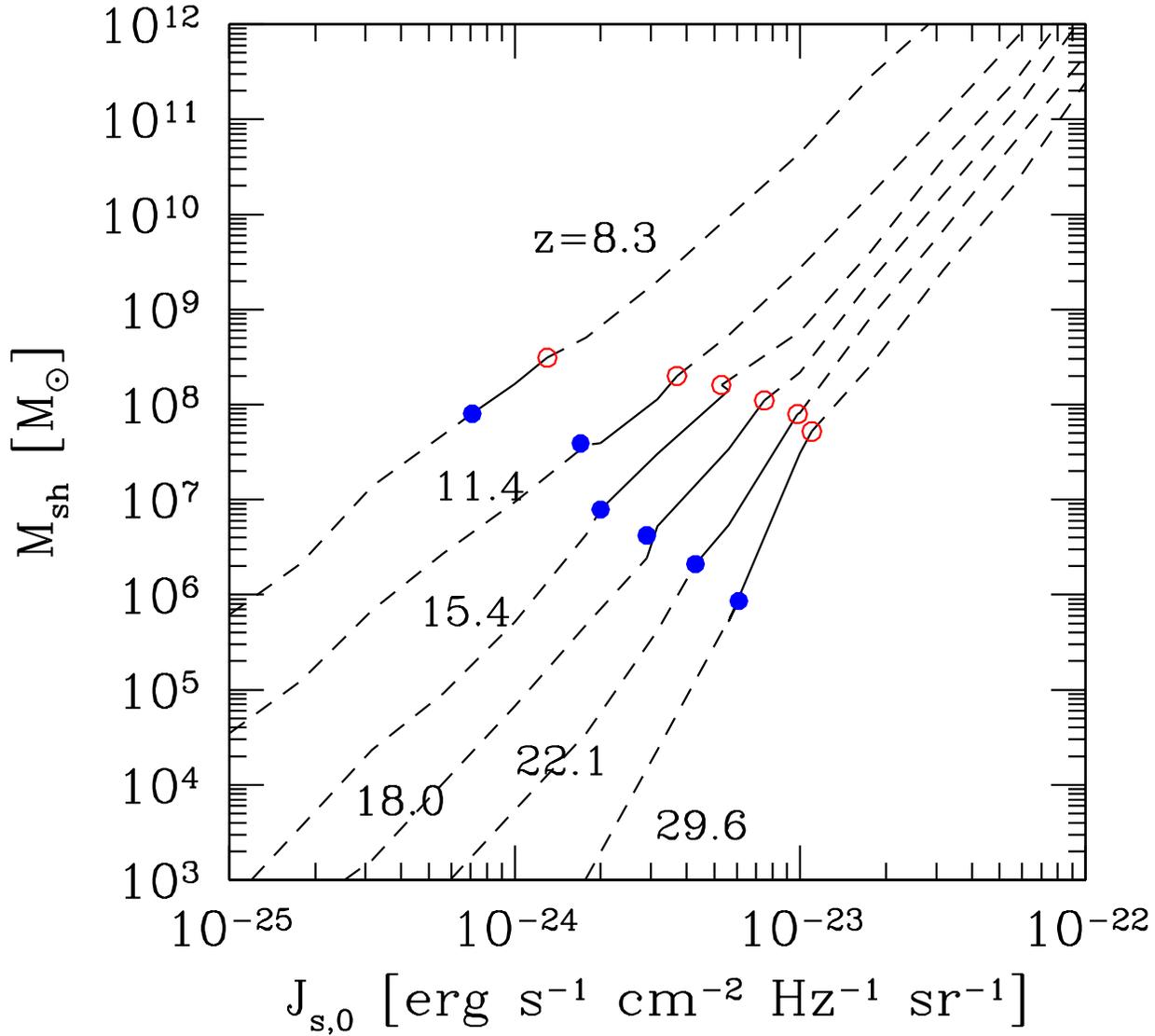}}
\caption{\label{fig5}{Minimum total mass for
self-shielding from an external
incident flux with intensity $J_{s,0}$ at the Lyman limit. The curves
are for different redshift: from the top to the bottom $z=$8.3, 11.4,
15.4, 18.0, 22.1, 29.6. Circles show the value of $M_H$ (open) and
$M_{crit}$ (filled). Radiative feedback works only in the solid portions
of the curves. }}
\end{figure}

\begin{figure}[t]
\centerline{\psfig{figure=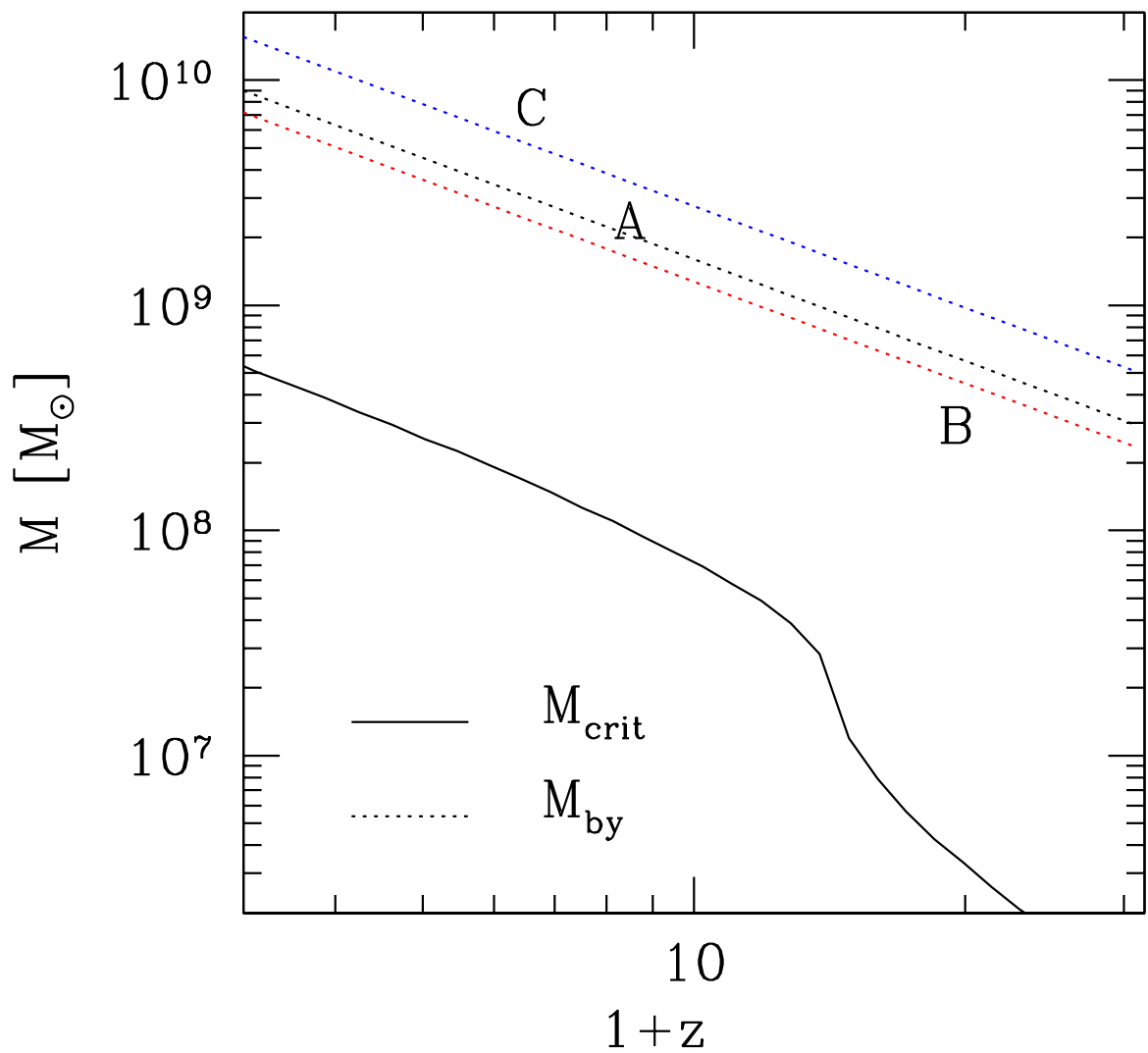}}
\caption{\label{fig6}{Minimum total mass for collapse of an object            
as a function of redshift according to Tegmark \etal (1997)
($M_{crit}$, solid line) and maximum mass 
for blowaway as a function of redshift ($M_{by}$,
dotted line). $M_{by}$ is shown for the three runs A, B, C (see Tab.
1).  }}
\end{figure}

\begin{figure}[t]
\centerline{\psfig{figure=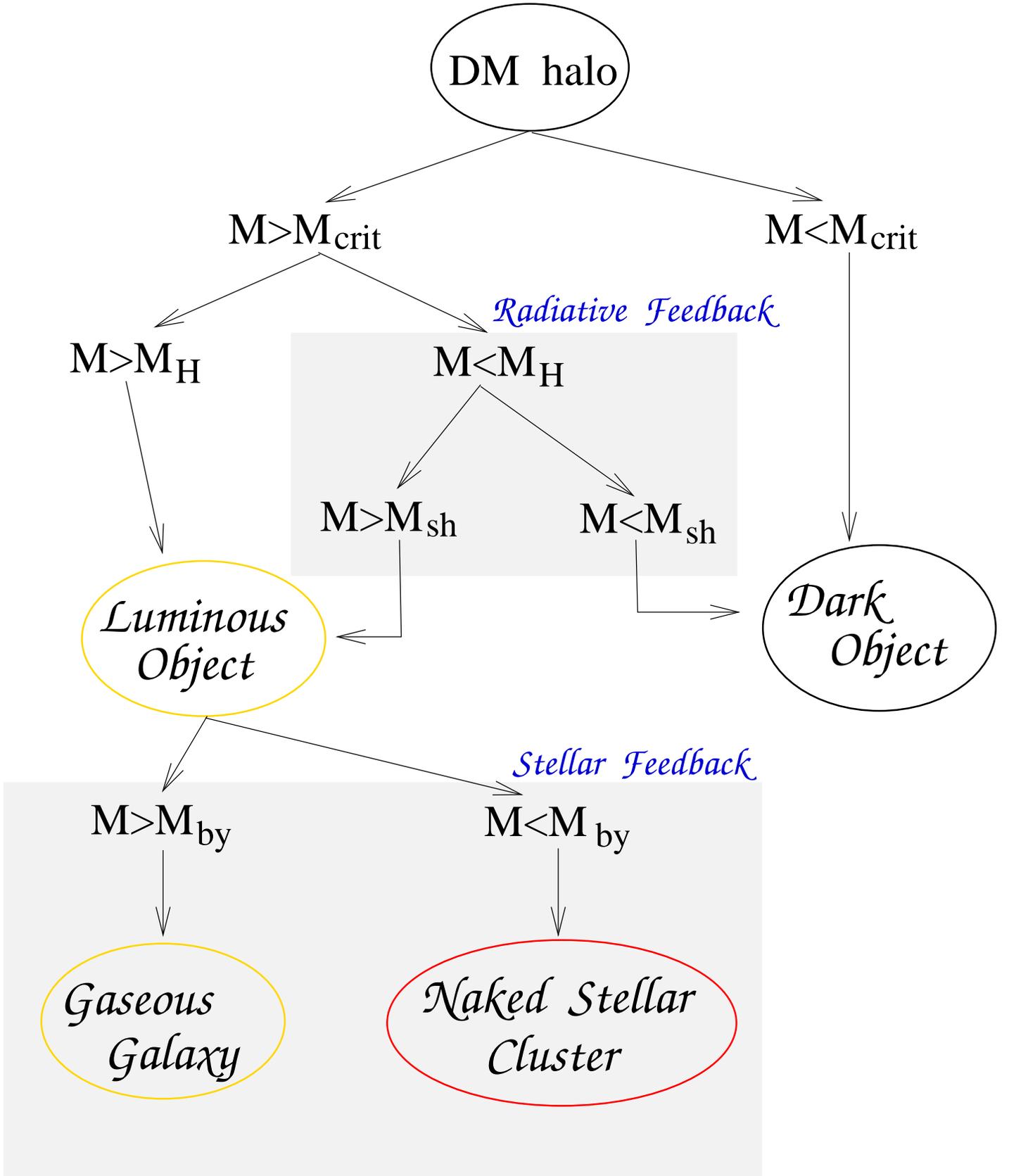}}
\caption{\label{fig7}{Possible evolutionary tracks of objects as
determined by the processes and feedbacks included in the model.
Explanations are given in \S 6.}}
\end{figure}

\begin{figure}[t]
\centerline{\psfig{figure=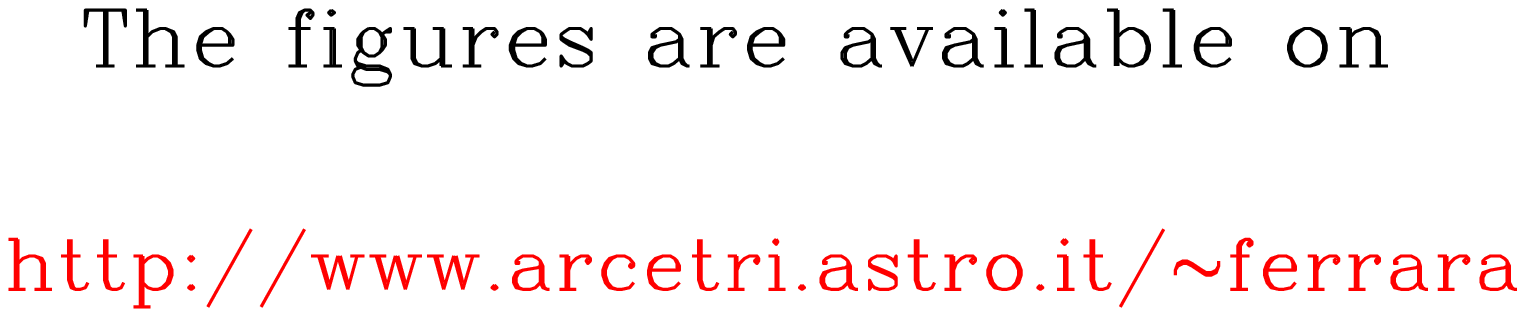}}
\caption{\label{fig8}{Global spatial structure of photoionized regions   
at $z$=22.1 (a), 19.8 (b), 18.0 (c)  and 15.4 (d).}}
\end{figure}

\begin{figure}[t]
\centerline{\psfig{figure=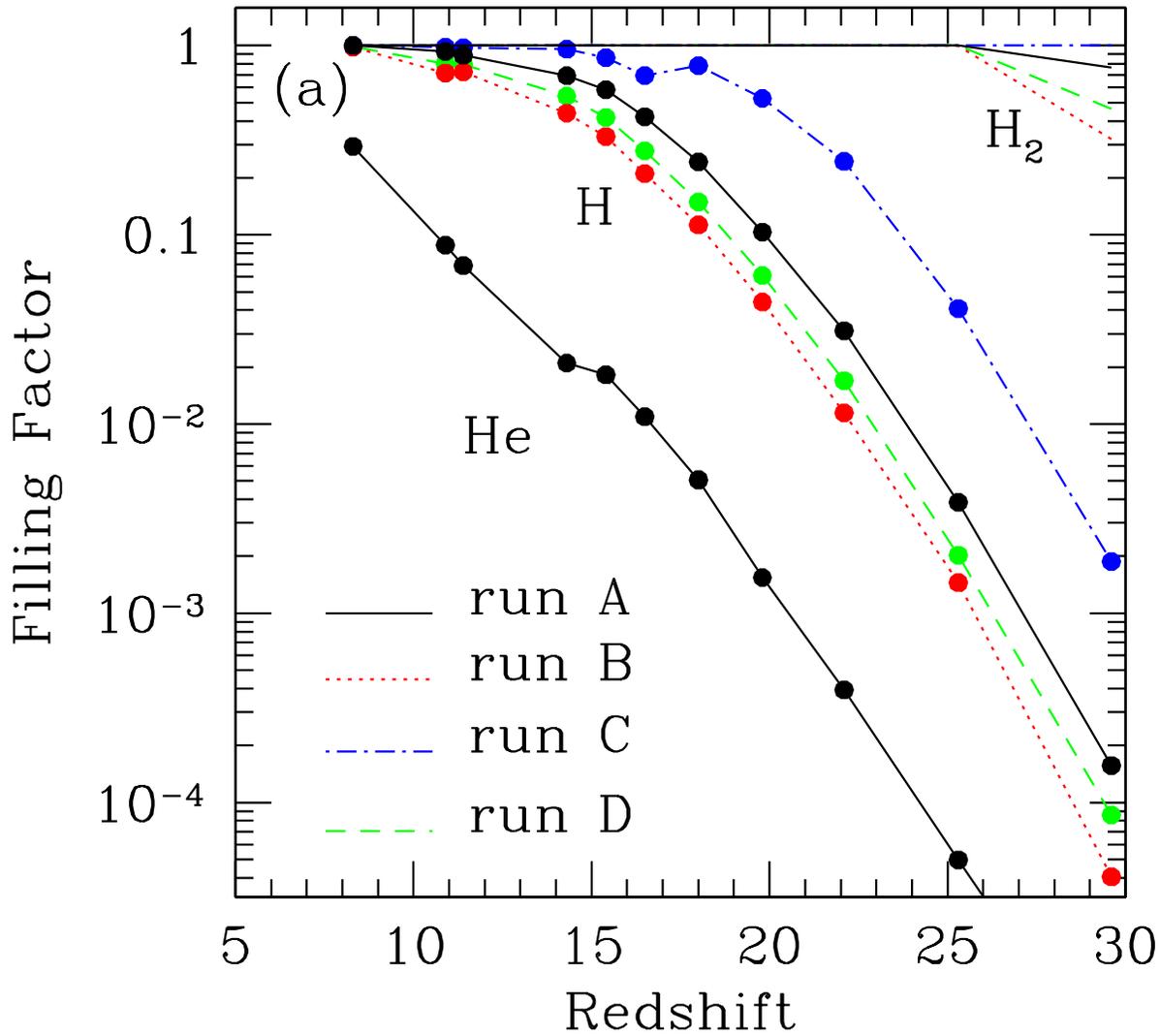}}
\caption{\label{fig9}{(a) dissociated molecular hydrogen (upper set of lines),
ionized atomic hydrogen (middle set) and doubly ionized helium
(bottom line)  filling factor as a function of redshift for different runs:
A (solid line), B (dotted), C (dashed-dotted) and D (dashed).
(b) same as (a) for runs: A (solid line), A1 (dotted)
and A2 (dashed-dotted).}}
\end{figure}

\begin{figure}[t]
\centerline{\psfig{figure=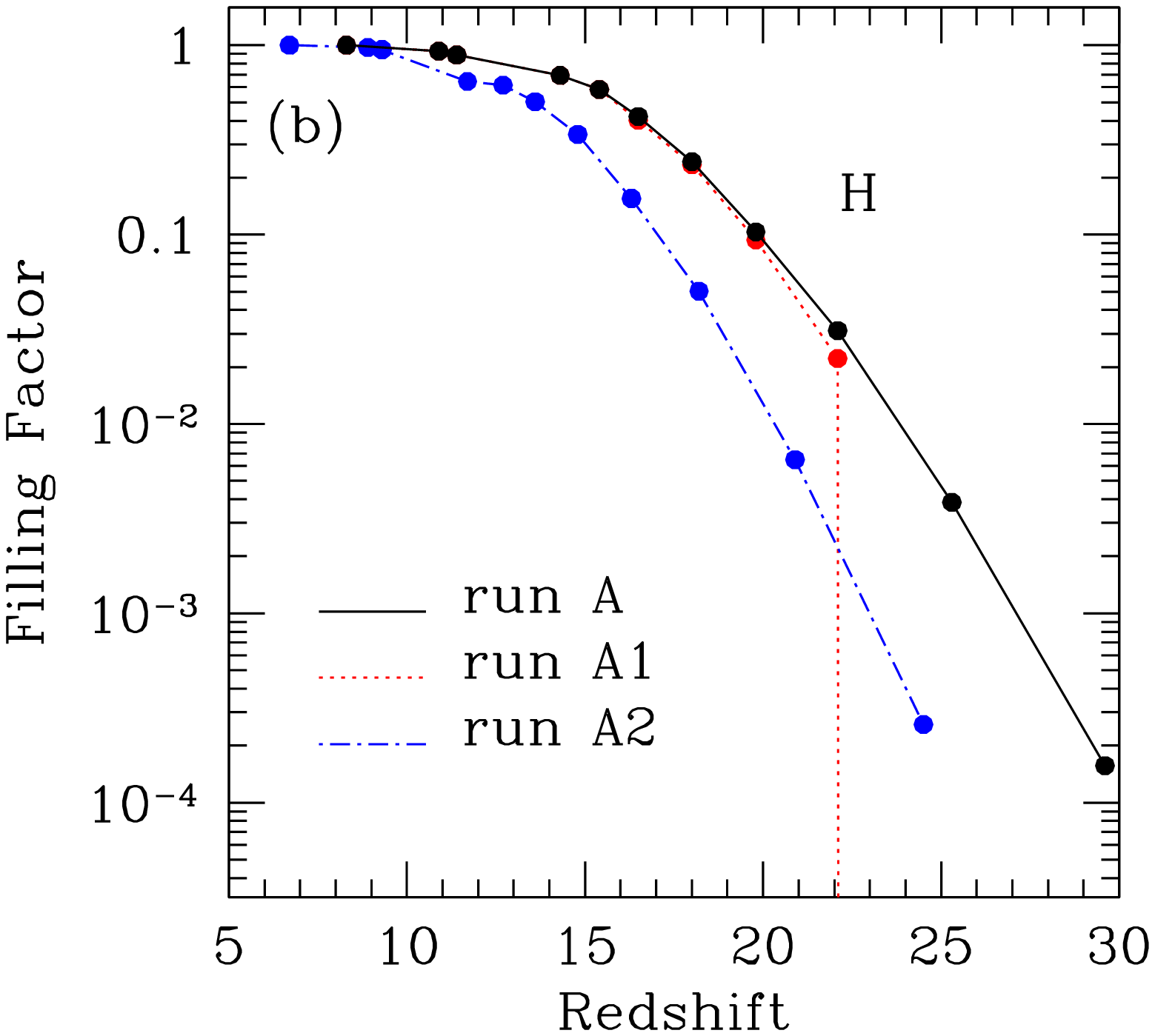}}
\end{figure}

\begin{figure}[t]
\centerline{\psfig{figure=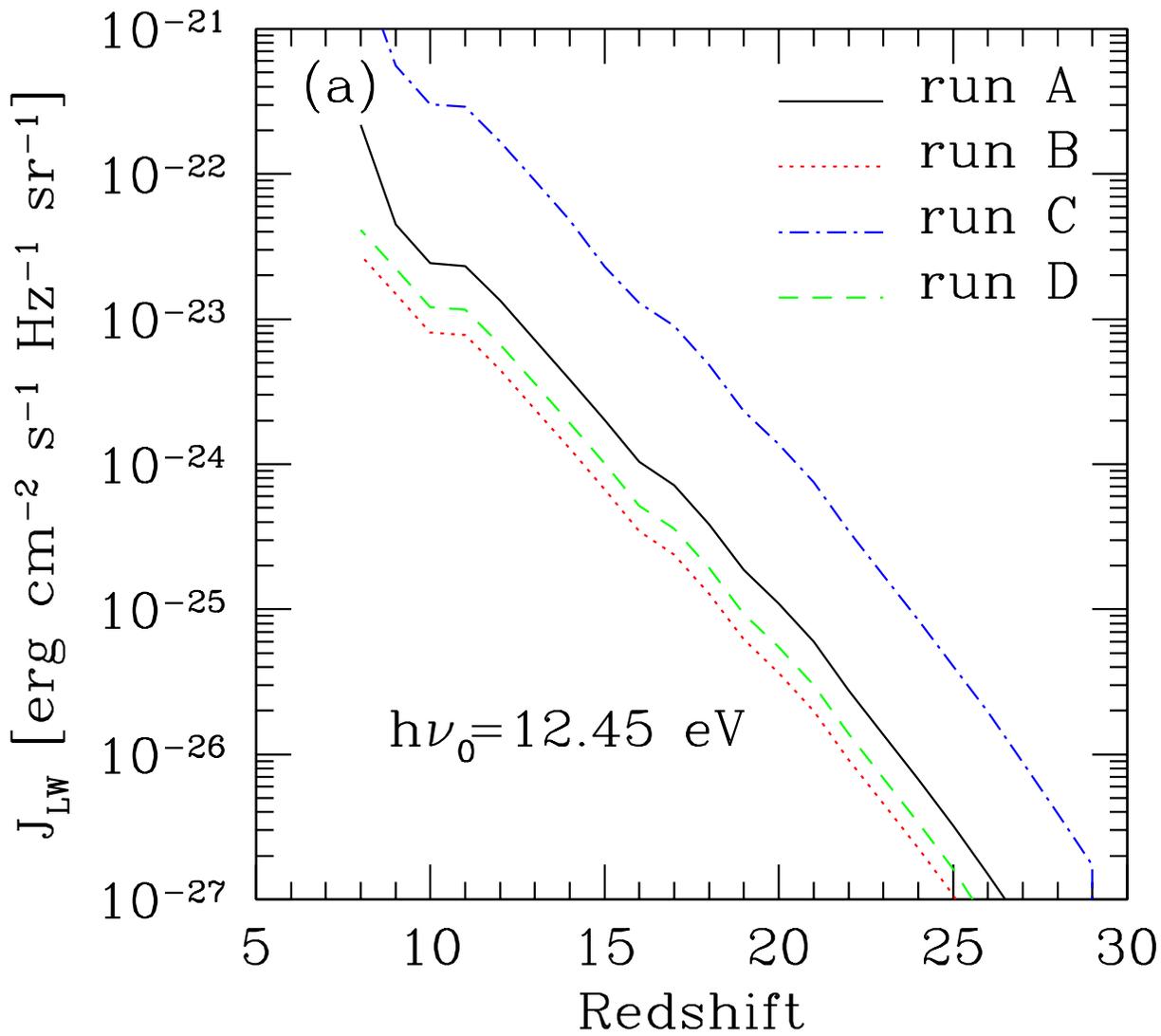}}
\caption{\label{fig10}{(a) SUVB intensity as a function of redshift, at the
central frequency of the Lyman-Werner band $h\nu_0$=12.45 eV for runs A-D. 
(b) same as (a) for runs A, A1, A2. 
}}
\end{figure}

\begin{figure}[t]
\centerline{\psfig{figure=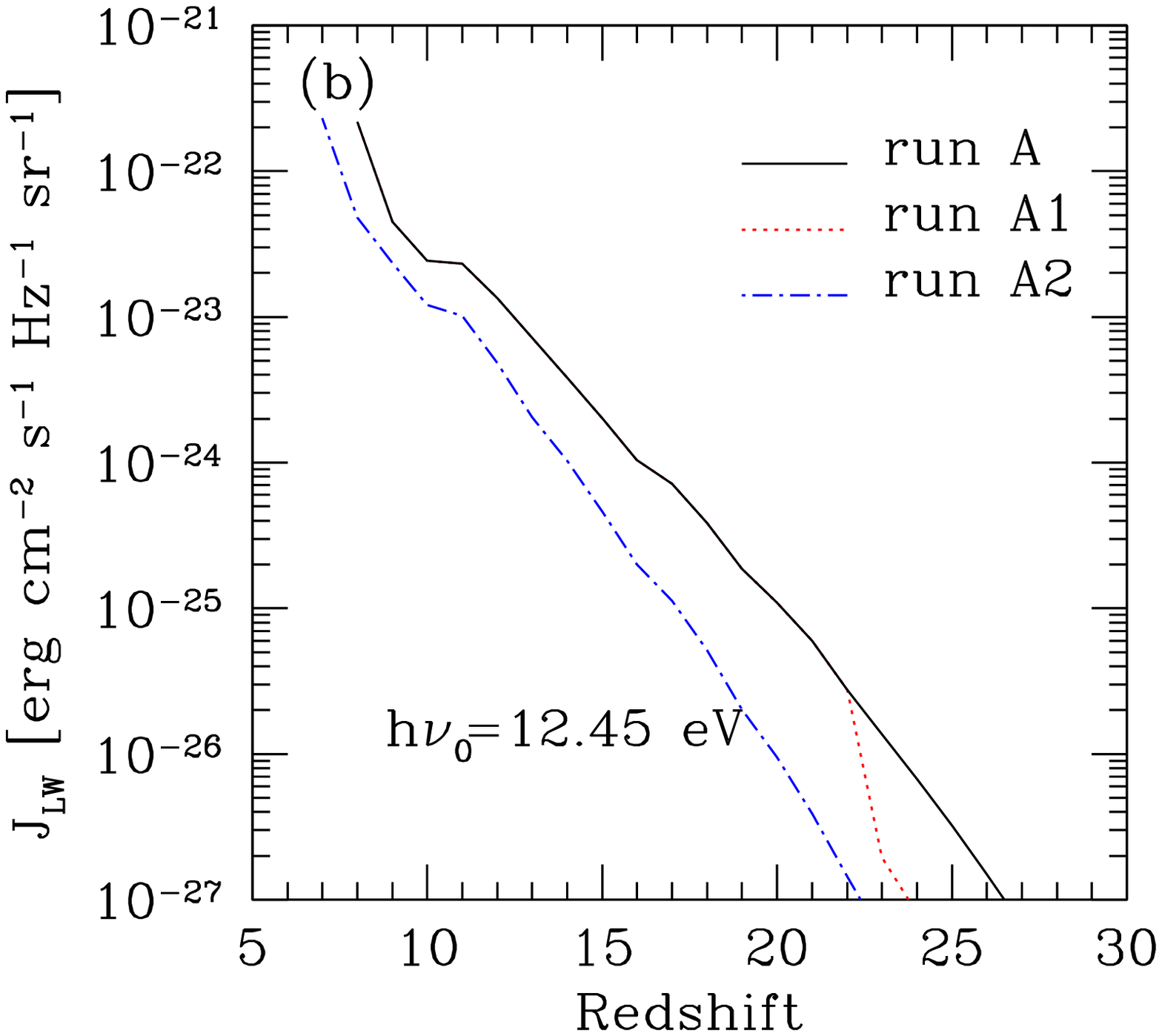}}
\end{figure}

\begin{figure}[t]
\centerline{\psfig{figure=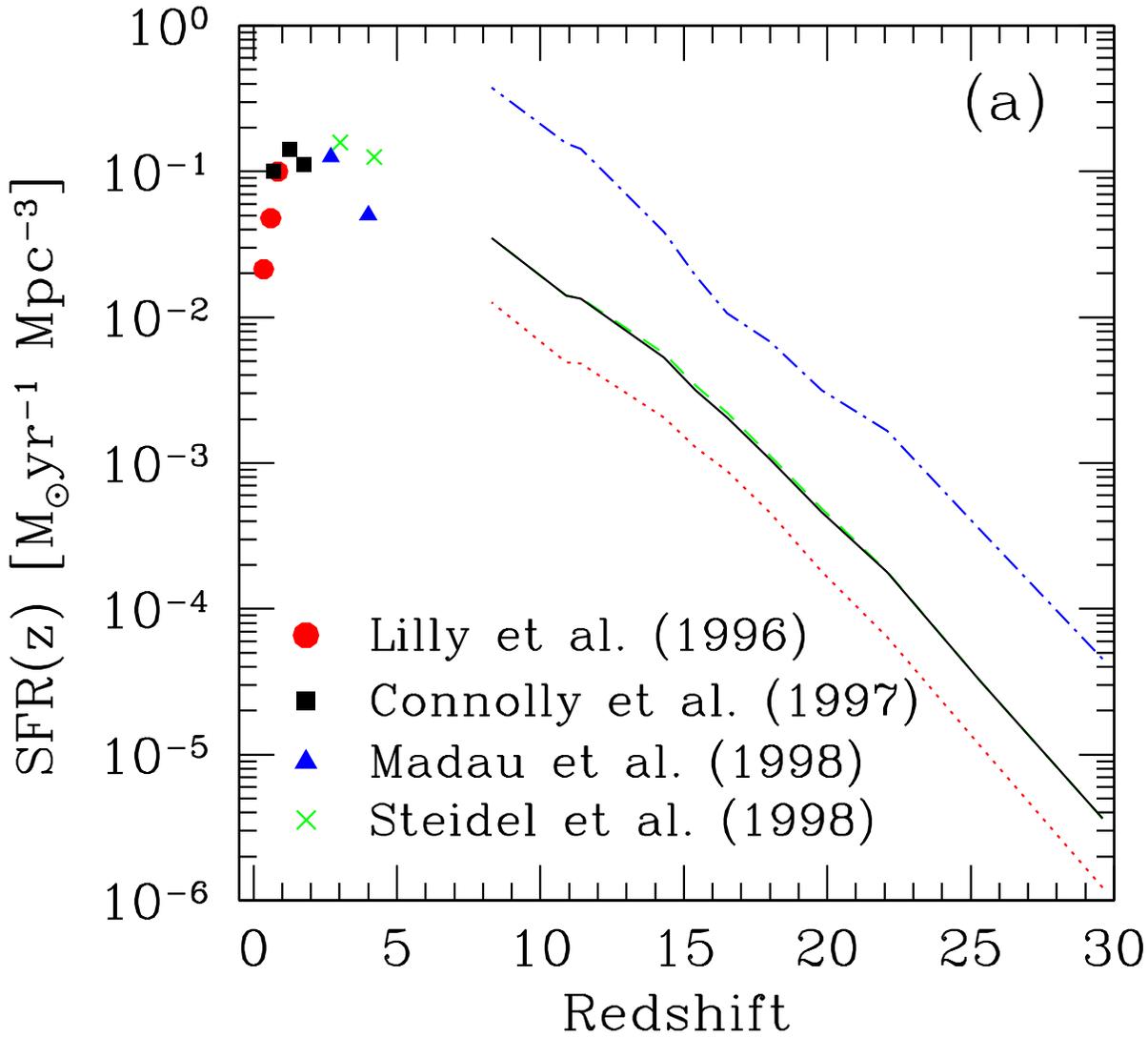}}
\caption{\label{fig11}{(a) comoving star formation rate as a
function of redshift, for runs C, D, A, B, from the top to the bottom,
respectively. Data              points are from
Lilly \etal (1996) [circles]; Connolly et
al. (1997) [squares]; Madau, Pozzetti \& Dickinson (1998) 
[triangles] and Steidel \etal  (1998) [crosses]. 
(b) same as (a) for runs A, A1, A2.}} 
\end{figure}

\begin{figure}[t]
\centerline{\psfig{figure=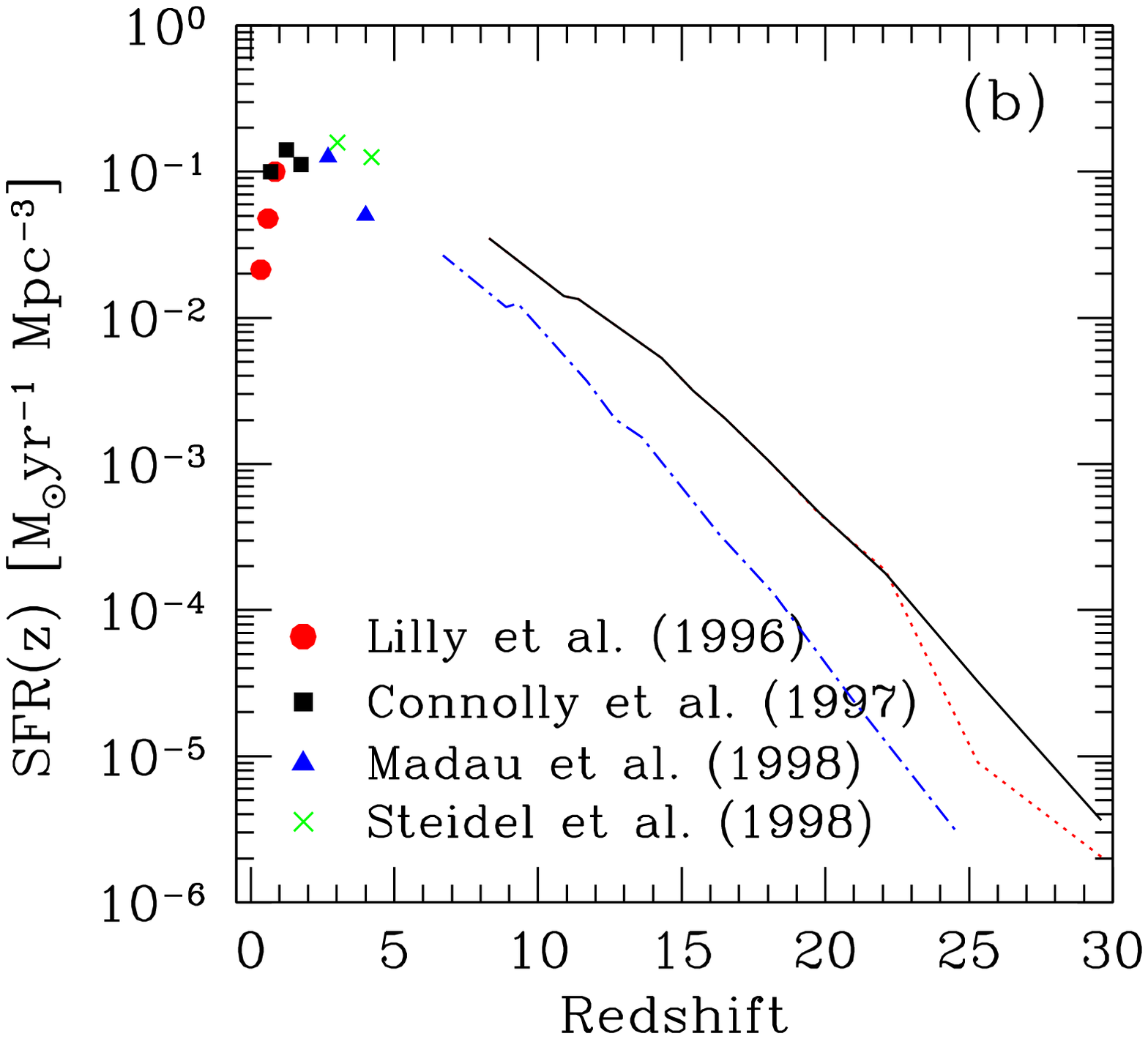}}
\end{figure}

\begin{figure}[t]
\centerline{\psfig{figure=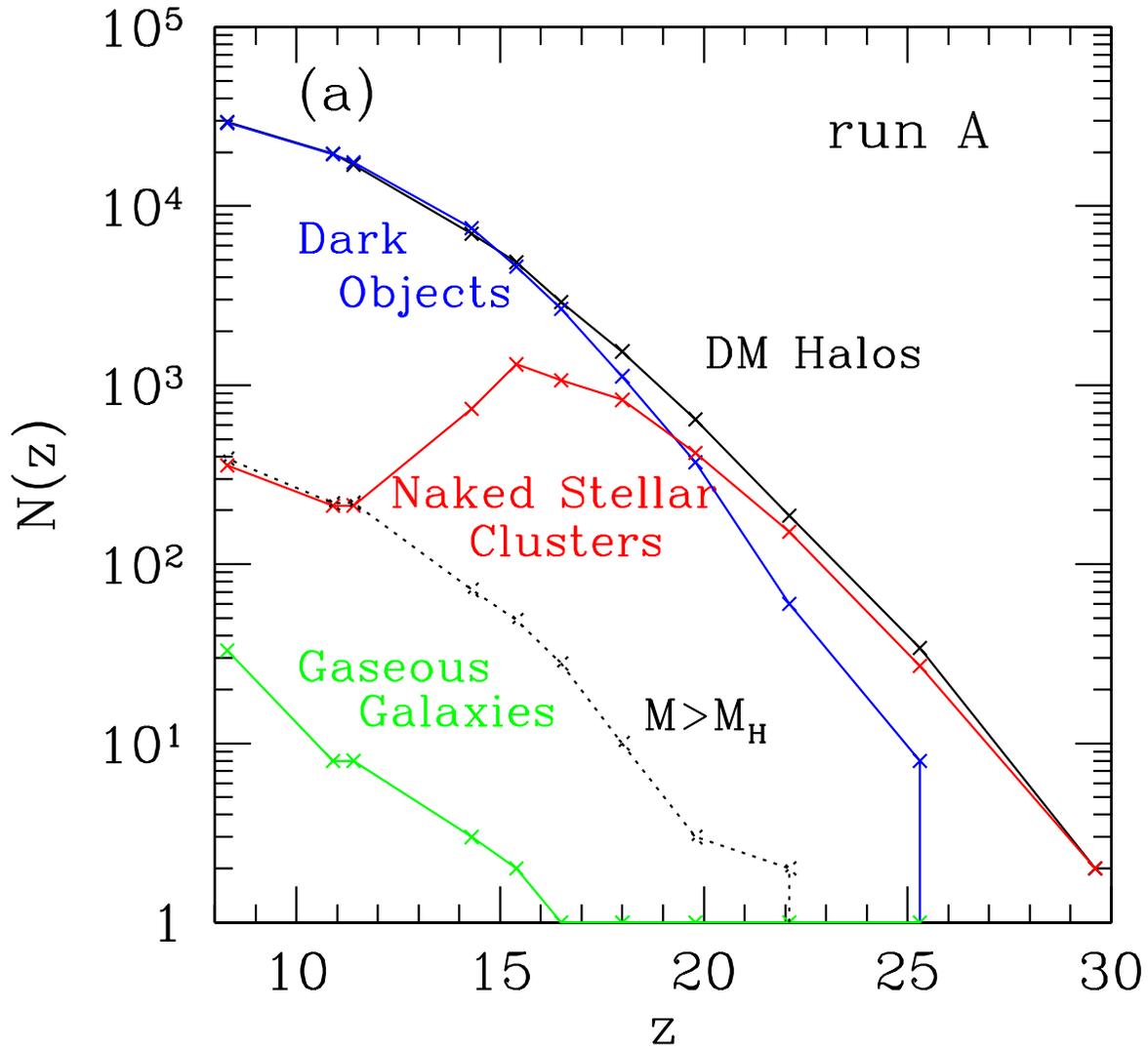}}
\caption{\label{fig12}{(a) Number evolution of different objects in the
simulation box for run A (see discussion in \S~7.4) 
(b) same as (a) for run C.}}
\end{figure}

\begin{figure}[t]
\centerline{\psfig{figure=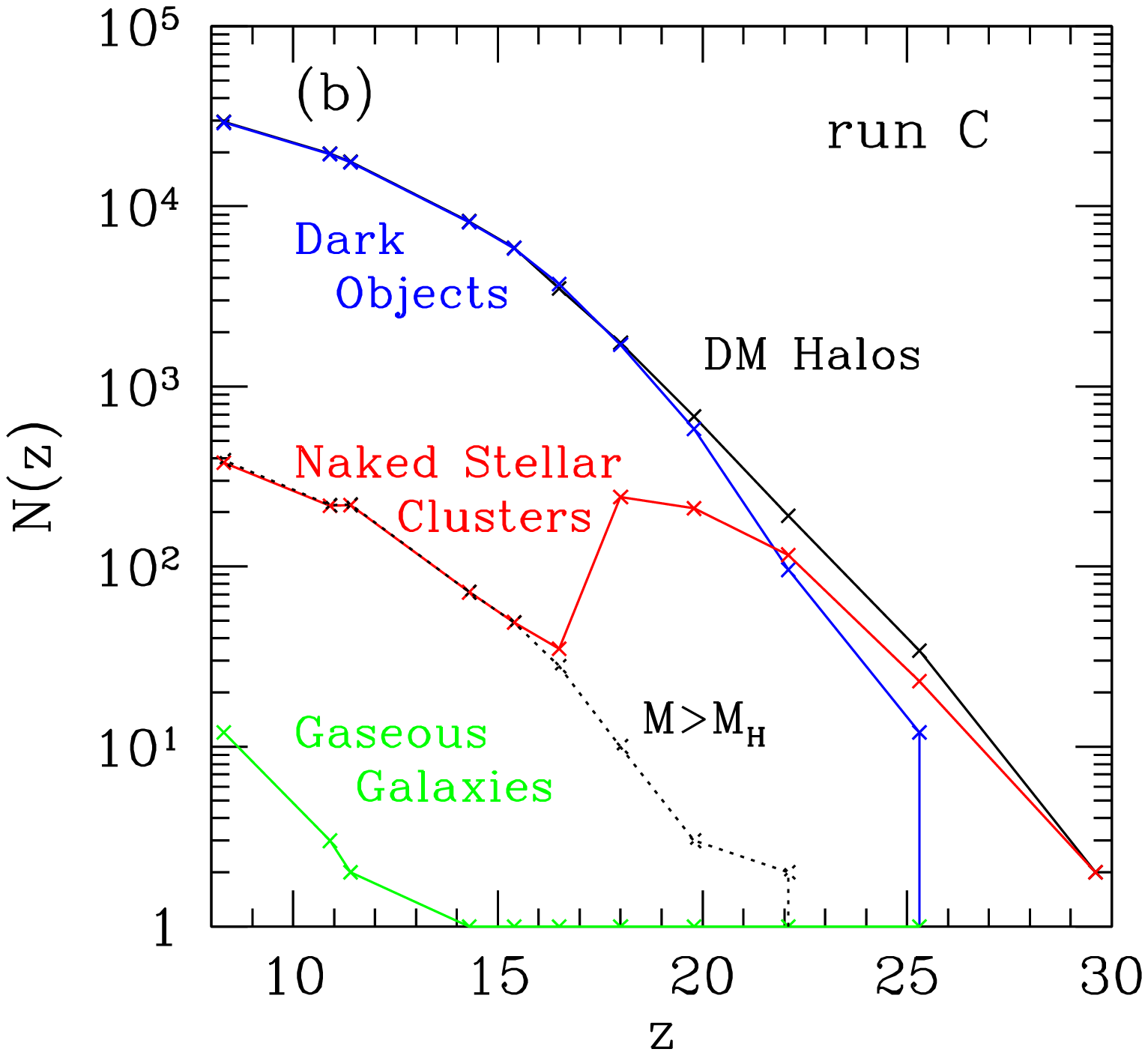}}
\end{figure}

\begin{figure}[t]
\centerline{\psfig{figure=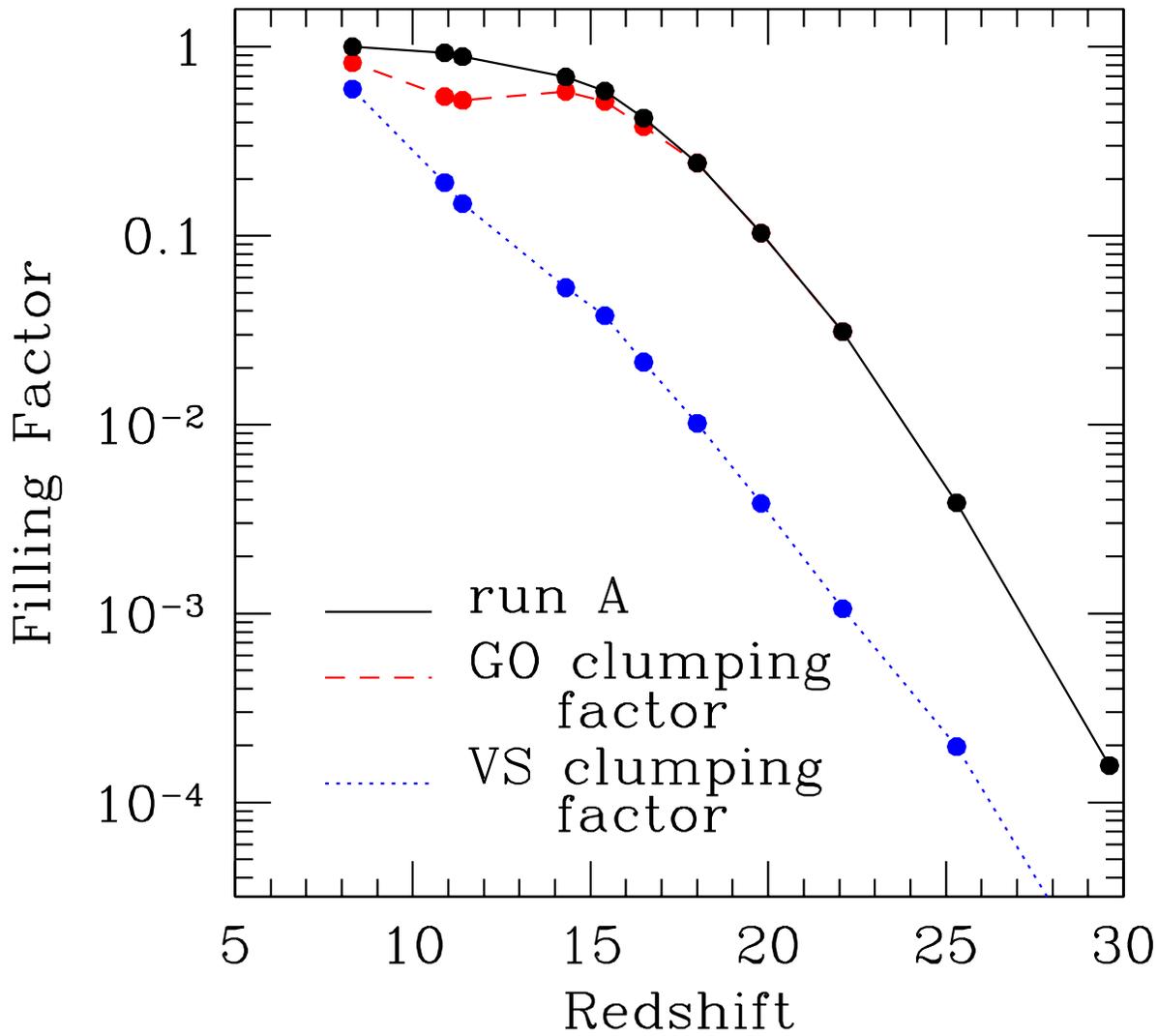}}
\caption{\label{fig13}{Ionized atomic hydrogen filling factor as a function 
of redshift for different runs: A (solid line), A with the clumping
factor curve of GO (dashed) and VS (dotted).}}
\end{figure}

\vfill
\eject
\end{document}